\documentclass[5p,twocolumn,10pt,times]{elsarticle}
\usepackage{amsmath}
\usepackage{hyperref}
\usepackage{graphicx}
\usepackage{subfigure}
\usepackage{verbatim}
\usepackage{ulem}
\usepackage[ruled]{algorithm2e} 
\usepackage{algorithmic}
\usepackage{multirow}
\usepackage{caption}
\usepackage{amsmath}
\usepackage{ctable}
\biboptions{sort&compress}
\graphicspath{{images/}}
\usepackage{color}

\newcommand{\rev}[2]{#2}
\newcommand{\revnew}[2]{#2}

\begin{document}

\begin{frontmatter}

\title{Computer-Controlled 3D Freeform Surface Weaving}

\author[1,2]{Xiangjia Chen}
\author[2]{Lip M. Lai}
\author[2]{Zishun Liu}
\author[2]{Chengkai Dai}
\author[2]{Isaac C.W. Leung}
\author[3]{Charlie C.L. Wang\corref{mycorrespondingauthor}}
\author[1,2]{Yeung Yam\corref{mycorrespondingauthor}}
\cortext[mycorrespondingauthor]{Corresponding authors: Charlie C.L. Wang and Yeung Yam (E-mail: changling.wang@manchester.ac.uk; yyam@mae.cuhk.edu.hk)}

\address[1]{Department of Mechanical and Automation Engineering, The Chinese University of Hong Kong, Shatin, Hong Kong, China}
\address[2]{Centre for Perceptual and Interactive Intelligence (CPII) Limited, Hong Kong, China}
\address[3]{Digital Manufacturing Lab, School of Engineering, The University of Manchester, United Kingdom}

\begin{abstract} 
In this paper, we present a new computer-controlled weaving technology that enables the fabrication of woven structures in the shape of given 3D surfaces by using threads in non-traditional materials with high bending-stiffness, allowing for multiple applications with the resultant woven fabrics. A new weaving machine and a new manufacturing process are developed to realize the function of 3D surface weaving by the principle of short-row shaping. A computational solution is investigated to convert input 3D freeform surfaces into the corresponding weaving operations (indicated as W-code) to guide the operation of this system. A variety of examples using cotton threads, conductive threads and optical fibres are fabricated by our prototype system to demonstrate its functionality. 

\end{abstract}


\end{frontmatter}

\section{Introduction}\label{secIntroduction}
The history of textile fabrication is probably as old as human history\cite{adanur2020handbook}. With the development of civilization, the function of textile products has been extended from keeping warmth to smart wearable devices in applications such as aerospace engineering \cite{kelkar2006structural} and biomedical engineering \cite{di2006magic,9852702}. There are growing demands for 3D shaping new composite materials~\cite{mouritz1999review} and e-textile~\cite{komolafe2021textile,levitt20203d}, where the geometry of fabrics could possibly be formed by four different methods including knitting~\cite{mccann2016compiler,narayanan2018automatic,liu2019computational,igarashi2008knitting}, weaving~\cite{bilisik2010multiaxis,wu2020automatic,ren20213d}, felting\cite{chen2015advances} and robotic deposition / winding ~\cite{shirinzadeh2004fabrication,shirinzadeh2007trajectory,SORRENTINO201743}. 

\begin{figure}[t]
\centering 
\includegraphics[width=\linewidth]{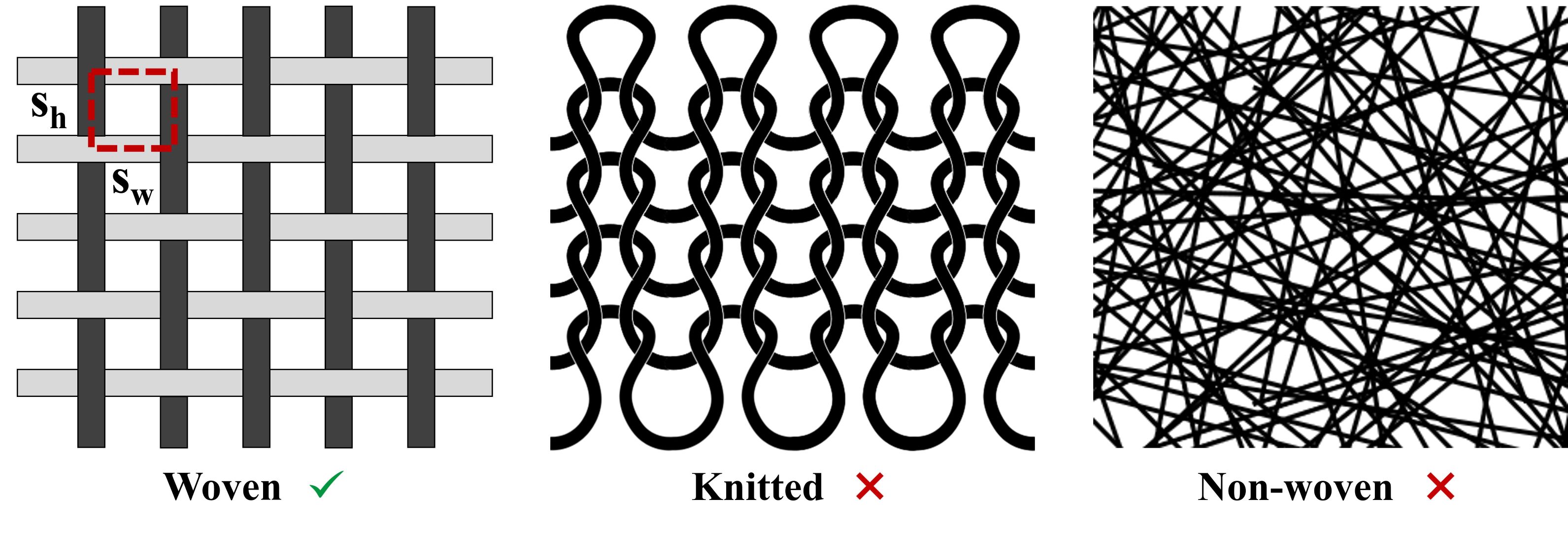}\vspace{-10pt}
\caption{\rev{The system supports the fabrication}{An illustrations} of woven structures using non-traditional materials with high bending-stiffness, which can neither be realized by the knitting process (i.e., highly bent loops are formed) nor the felting process (i.e., none woven structure is formed). The width $s_w$ of a woven structure is determined by the mechanical components as reed and the height $s_h$ can be controlled by the motors \& gears of warp beams.}\label{fig:stitch}
\end{figure}
\begin{figure}[t]
\centering 
\includegraphics[width=\linewidth]{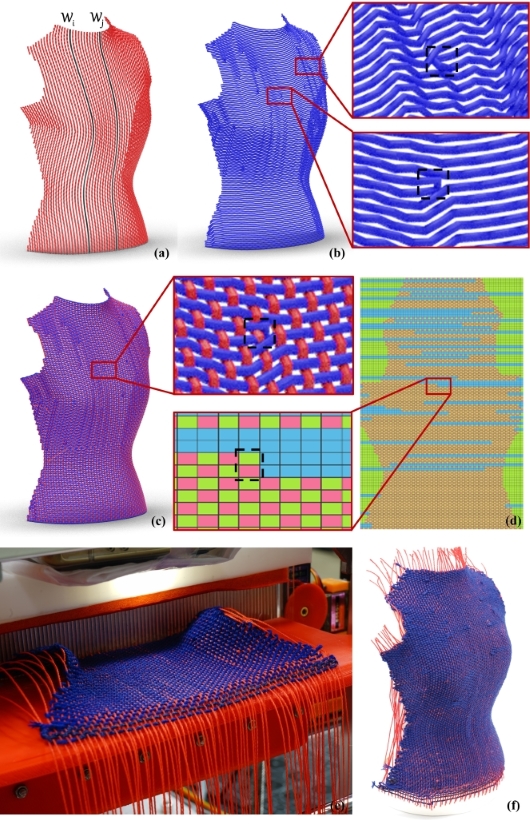}\vspace{-10pt}
\caption{Given the 3D surface of a vest's front piece\rev{~(a), an algorithm is developed to generate the weaving information (as the weaving map shown in (b)) for controlling the operations of our 3D surface weaving machine (c.1 \& c.2)}{, the 3D woven structures (c) are formed by controlling (a) the lengths of warp threads in different columns -- e.g., $W_i$ and $W_j$ have different lengths and (b) the different starting / ending positions of the weft thread to form short rows -- see the regions circled by dash lines}. 
\rev{}{Fabricating the 3D woven structures is supervised by a weaving map as shown in (d).}
The result of fabrication by cotton threads are shown in \rev{(d)}{(e, f)} with red for warp threads and blue for weft threads. }\label{fig:teaser}
\end{figure}

In the \rev{resultant fabrics of}{fabrics formed by} felting, threads / fibres are randomly distributed (see the \rev{lower }{}right of Fig.\ref{fig:stitch}) -- i.e., not controllable. The approaches of robotic direct deposition can only fabricate structures with weakly bonded threads / fibres -- i.e., the threads are `stitched together' by chemical but not mechanical bonding in a manner similar to filament based 3D printing~\cite{shirinzadeh2004fabrication,shirinzadeh2007trajectory}, and the winding technique is mainly limited to the surfaces obtained by revolution~\cite{SORRENTINO201743,li2021Winding}. To overcome these problems, a new technology needs to be developed for fabricating woven structures with controllable locations of threads on 3D surfaces. Note that, for the sake of explanation, we simply call each unit structure of woven fabrics as a \textit{stitch} in the rest of this paper. \rev{The bottom row of Fig.\ref{fig:overview}}{Fig.\ref{fig:stitch}} shows the stitch structures in weaving (left), in knitting (middle), and the unstructured threads in felting (right). Generally, all structured stitches need to be mapped into the elements of a two-dimensional grid layout to enable the manufacturing process. Such a map is named as \textit{weaving map} (see Fig.\ref{fig:teaser}(b) for an example). Weaving process results in interlaced warp and weft threads that are nearly perpendicular to each other~\cite{adanur2020handbook} \rev{}{-- fabrics can therefore be formed by threads with high bending-stiffness}. \rev{}{This enables many applications using threads with advanced physical functions such as carbon fibres, optical fibres, and conductive wires (ref.~\cite{mouritz1999review,komolafe2021textile,levitt20203d}).}

Three-dimensional freeform shapes can be formed in knitting by the operations to increase or decrease the number of stitches, which has been commonly adopted in hand knitting for simple shapes such as a rounded cap since the 16th century~\cite{spencer2001knitting}. Different from weaving, the inter-looped stitches in knitting are formed by warping a continuous filling thread throughout the whole fabric-forming process. More complicated geometry was considered in computer graphics by employing increase-decrease shaping for hand knitting \cite{igarashi2008knitting} or short-row shaping for machine knitting \cite{liu2021knitting}. Increase / decrease shaping changes the local width of the fabric by adding an extra loop in a row of stitches (or trapping a loop to more than one loop in the previous row). Short-row shaping changes the local height of the fabric by knitting partial rows in regions needing more stitches. Modern computational techniques have been developed to generate knitting code for realizing desired shapes on knitting machines, including the compiler~\cite{mccann2016compiler}, the visual programming tool~\cite{narayanan2019visual}, and the algorithm for controlling the distribution of elasticity~\cite{liu2021knitting}. However, highly curved loops formed in knitting make it challenging to use materials with high bending-stiffness (see the middle of \rev{bottom row in Fig.\ref{fig:overview}}{Fig.\ref{fig:stitch}}). An alternative manufacturing method is needed. 

\rev{In this paper, we propose a new computer-controlled weaving technique that enables the fabrication of woven fabrics as 3D freeform surfaces (see Fig.\ref{fig:overview_machine})}{We aim at developing a new technology for computer-controlled fabrication of 3D freeform shapes by woven structures}. Different from the existing works on weaving 3D solid geometry like height-fields (ref.~\cite{wu2020Woven3D}), the weaving technique developed in our work focuses on producing 3D surface geometry. 3D surface weaving products can be widely seen in our daily life (e.g., woven baskets and woven chairs), which however are mainly hand-made. Ren et al.\cite{ren20213d} presented an optimization-based approach to \rev{solve the inverse design problem for}{design} woven structures by an algorithm to compute the ribbon's planar geometry. However, the manufacturing process has not been considered. \rev{}{The challenges for realizing a computer-controlled 3D surface weaving process include:} 
\begin{itemize}
\item \rev{}{To form a 3D surface patch, different numbers of stitches should be produced in different rows on the resultant woven structures -- therefore, a \textbf{new manufacturing process} different from flat weaving needs to be developed.}

\item \rev{}{A \textbf{new machine} to realize the 3D weaving process needs to be developed to enable the length control of warp threads and weft threads in different columns and rows.}

\item \rev{}{A \textbf{new algorithm} is also demanded to convert an input 3D freeform surface into the machine instruction of weaving operations.}
\end{itemize}
\rev{}{To the best of our knowledge, these do not exist in literature although the demanded algorithm shares some similarity to our earlier work of knitting map generation (ref. \cite{liu2021knitting}).} \revnew{}{Considering the requirements that 1) the first row and the last row of weaving are located on the boundary of 3D surface patches and 2) only one weft thread is allowed to travel during the process, the 3D surface patch utilized in this paper should be an oriented 2-manifold with disk-like topology.}

\section{Weaving 3D Freeform Surface}\label{secSurfWeaving}
To enable the \rev{novel approach of}{new manufacturing process of} 3D surface weaving, \rev{not only}{both} the new hardware \rev{but also}{and} the new algorithm are developed in our work. \rev{}{Basically, a 3D surface of woven structures can be formed by controlling the lengths of warp threads in different columns and the different starting / ending positions of the weft thread as illustrated in Fig.\ref{fig:teaser}(a, b).}

\begin{figure}[t]
\centering 
\includegraphics[width=\linewidth]{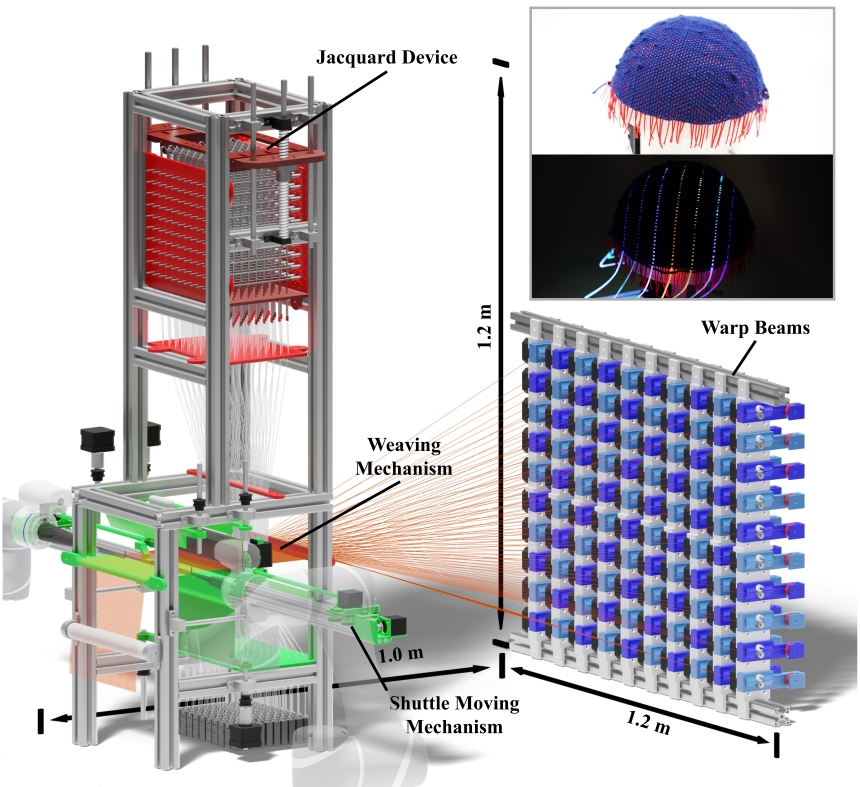}\vspace{-10pt}
\caption{Our \rev{3D surface}{computer-controlled 3D} weaving system can fabricate woven structures to form a target 3D surface shape and it mainly consists of \rev{three}{four} hardware parts: 1) jacquard device, 2) warp beams, 3) weaving mechanism\rev{}{, and 4) shuttle moving mechanism}. \rev{}{The weaving process can be completed automatically with the help of two robotic arms equipped with camera \revnew{}{together with a camera mounted above the weaving region}.} The weaving result as a hemisphere with embedded optical fibres is shown in the right-upper corner.}\label{fig:overview_machine}
\end{figure}

Jacquard device\cite{miura1964jacquard} on a flat weaving machine\rev{}{, which is originally used to product flat embossed patterns on fabrics,} is first modified to enable the function of partial weaving by controlling different up and down states of warp threads (i.e., those along the column directions). As a result, different numbers of warp threads can be woven by the weft thread (i.e., along the row direction) starting from / ending at different locations -- see Fig.\ref{fig:teaser}\rev{(c)}{(b, c)} for an example and Sec.~\ref{subsecJacquardDevice} for more details. 
Secondly, considering that the shape of 3D woven fabric is determined according to the lengths of the warp threads in different columns, different warp threads need to be controlled to `tighten' or `release' individually (Sec.~\ref{subsecWarpBeams}). A \rev{}{new mechanism as a} matrix of warp beams is developed to realize this function (see the right of \rev{Fig.\ref{fig:overview}}{Fig.\ref{fig:overview_machine}}). How to effectively form the desired pulling operation while keeping the warp threads tight all the time during the weaving process is \rev{}{one of} the most challenging \rev{part}{tasks}. 
A weaving mechanism is designed for this purpose (Sec.~\ref{subsecWeavingMech}).  \rev{}{The other challenging task is weft length control, which is realized in our system by dual robotic-arms equipped with camera \revnew{}{together with a camera mounted above the weaving region} (Sec.~\ref{subsecWeftLenControl})}.

In the software aspect, an algorithm is developed to generate weaving maps for controlling the operations of \rev{3D surface weaving}{our weaving machine} (see Fig.\rev{\ref{fig:teaser}}{\ref{fig:teaser}(d)} for an example). The algorithm is based on first generating a knittable stitch mesh by using the geodesic field based method in \cite{liu2021knitting}. The stitch mesh for knitting is converted into a weaving map by considering the jacquard device and the weaving mechanism employed in our system. Each weaving map is later converted into a set of operations defined as W-code to govern the machine weaving. Details of the algorithm can be found in Sec.~\ref{subsecWCode}. 

In summary, the \textit{technical contributions} of our work include:
\begin{itemize}
\item A new \rev{method}{manufacturing process} for three-dimensional surface weaving that can employ warp / weft threads with high bending-stiffness;

\item The novel design of a computer-controlled machine for 3D surface weaving \rev{}{that enables the length control of warp and weft threads in different columns and rows};

\item A computational solution that can convert an input 3D freeform surface into the corresponding weaving operations.
\end{itemize}
\rev{To the best of our knowledge, automatic 3D surface weaving approach does not exist in the literature.}{} 
The performance of our technology has been tested and verified on a variety of examples in 3D freeform shape. Its potential applications will be demonstrated by woven surfaces fabricated by our prototype implementation with embedded conductive threads and optical fibres.

\section{\rev{}{Process and }Hardware Design}\label{secHardware}
We present \rev{}{the manufacturing process and} the hardware design of our 3D surface weaving system in this section. The following design parameters are adopted for the prototype of our machine to aim at the proof-of-concept applications. The dimensions of our machine are \rev{1.0m (length) $\times$ 0.8m (width) $\times$ 1.2m (height)}{1.0m (length) $\times$ 1.2m (width) $\times$ 1.2m (height)} -- see \rev{Fig.\ref{fig:overview}}{Fig.\ref{fig:overview_machine}}. The maximal width of fabrics supported by our machine is designed as 198mm, which is formed by 100 warp threads with a fixed stitch width $s_w$ as 2mm. The length of woven fabrics is theoretically unlimited, and the stitch height $s_h$ can be controlled by the warp beams in our design. The minimal stitch height is determined according to the material and the dimension of the weft threads used in weaving.

\begin{figure}[t]
\centering 
\includegraphics[width=\linewidth]{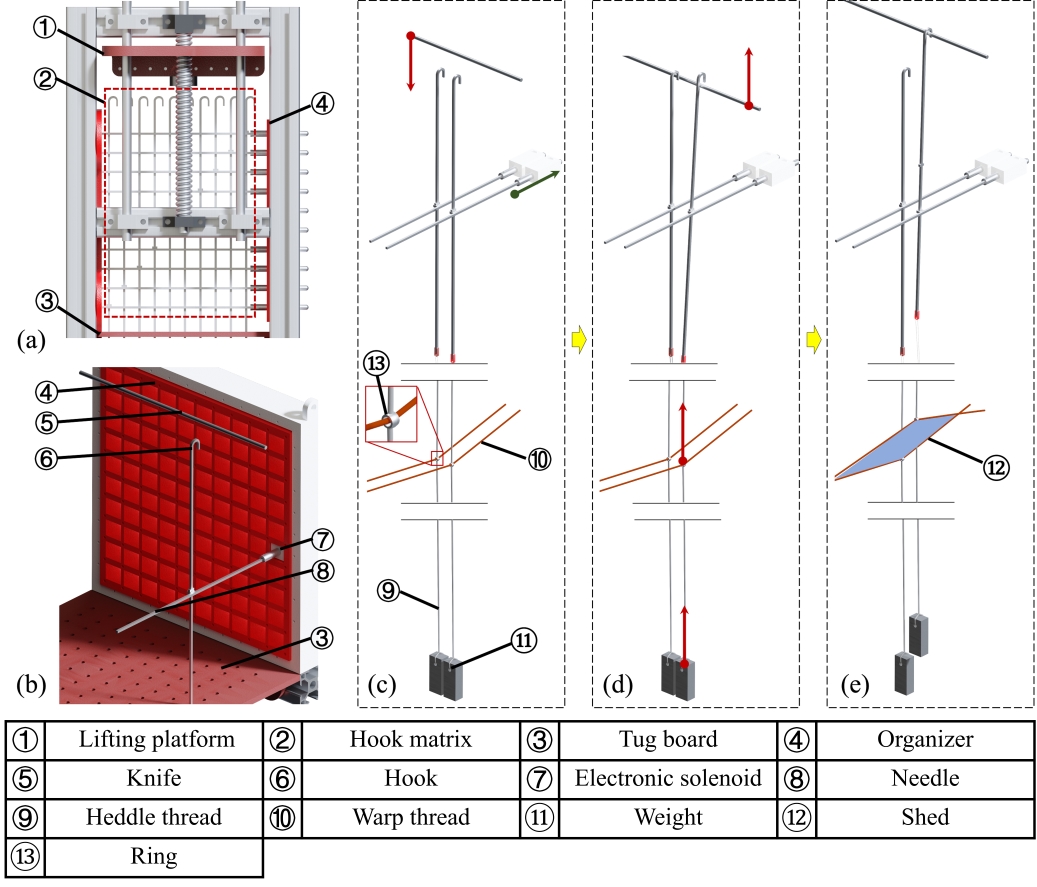}\vspace{-10pt}
\caption{\rev{}{The jacquard device (a, b) contains 13 components to lift the selected warp threads under computer control. The parsing motions of two heddle lifters with dropping and lifting commands are shown as: (c) the initial state, (d) the hanging-up state, and (e) the lifted state.}}\label{fig:jacquard_revise}
\end{figure}

\subsection{\rev{}{Partial weaving by} Jacquard device}\label{subsecJacquardDevice}
\rev{}{One major idea for forming the woven structure on a 3D freeform surface is to `partially' form stitches on different rows during the weaving process. Woven structures are formed by lifting up and pulling down warp threads in an interlaced way. Specifically, woven stitches will only be formed in the region where different up / down statuses are given to the neighboring warp threads so that they are inter-weaved with the weft thread. When neighboring warp threads are in the same statue of up (or down), the weft thread will bypass them to `skip' the stitch. Stitches are selectively added into the resultant fabrics by this partial weaving process -- therefore 3D surface can be form in a way similar to the short-row shaping technique in knitting.}

A jacquard device is employed in our system to selectively lift up warp threads to form a shed so that the shuttle holding the weft thread can travel through. Such a device is commonly found in traditional flat weaving machines to realize embossed patterns on woven fabrics. Differently, a jacquard device is introduced in our system to realize \rev{short-row shaping on a weaving machine}{partial weaving on our machine}. \rev{Specifically, woven stitches will only be formed in the region where different up / down statuses are given to the neighboring warp threads so that they are inter-weaved with the weft thread.}{} As illustrated in \rev{Fig.\ref{fig:jacquard}}{Fig.\ref{fig:jacquard_revise}}, a lifting platform with 10 knives is driven by a pair of screws to move up and down to lift (or drop) the selected heddle threads synchronously (the red arrows in \rev{Fig.\ref{fig:jacquard}}{Fig.\ref{fig:jacquard_revise}}(c) and (d)). Each hook can be individually controlled by a push or a pull motion (as indicated by the green arrow in \rev{Fig.\ref{fig:jacquard}}{Fig.\ref{fig:jacquard_revise}}(c)) with the help of a needle connected to an electronic solenoid. This controls whether a hook will be hung on the corresponding knife to move together. The solenoid can become shortened when receiving a control command. As a result, the attached needle will position the corresponding hook to the location to be lifted by a knife.

The parsing motions are given in \rev{Fig.\ref{fig:jacquard}}{Fig.\ref{fig:jacquard_revise}}(c)-(e) to illustrate how to form a shed by two hooks holding two neighboring warp threads. In this example, the right hook is selected to lift its corresponding warp thread while the left hook keeps dropping. First of all, the hooks are all hanging on their corresponding needles when the lifting platform is at a higher level in the initial state (see \rev{Fig.\ref{fig:jacquard}}{Fig.\ref{fig:jacquard_revise}}(c)). When a command is received, the lifting platform moves down to the height where the hooks can hang over the knives by shortening the solenoids to move the needles (as indicated by the green arrow in \rev{Fig.\ref{fig:jacquard}}{Fig.\ref{fig:jacquard_revise}}(c)). After pulling the right hook to the hanging position (as \rev{Fig.\ref{fig:jacquard}}{Fig.\ref{fig:jacquard_revise}}(d)), the lifting platform moves up. As a result, the knife has lifted the right hook together with the warp thread. A shed is formed by the lifted warp thread and the warp thread held by the left hook (see \rev{Fig.\ref{fig:jacquard}}{Fig.\ref{fig:jacquard_revise}}(e)). 


\begin{figure}[t]
\centering 
\includegraphics[width=1.0\linewidth]{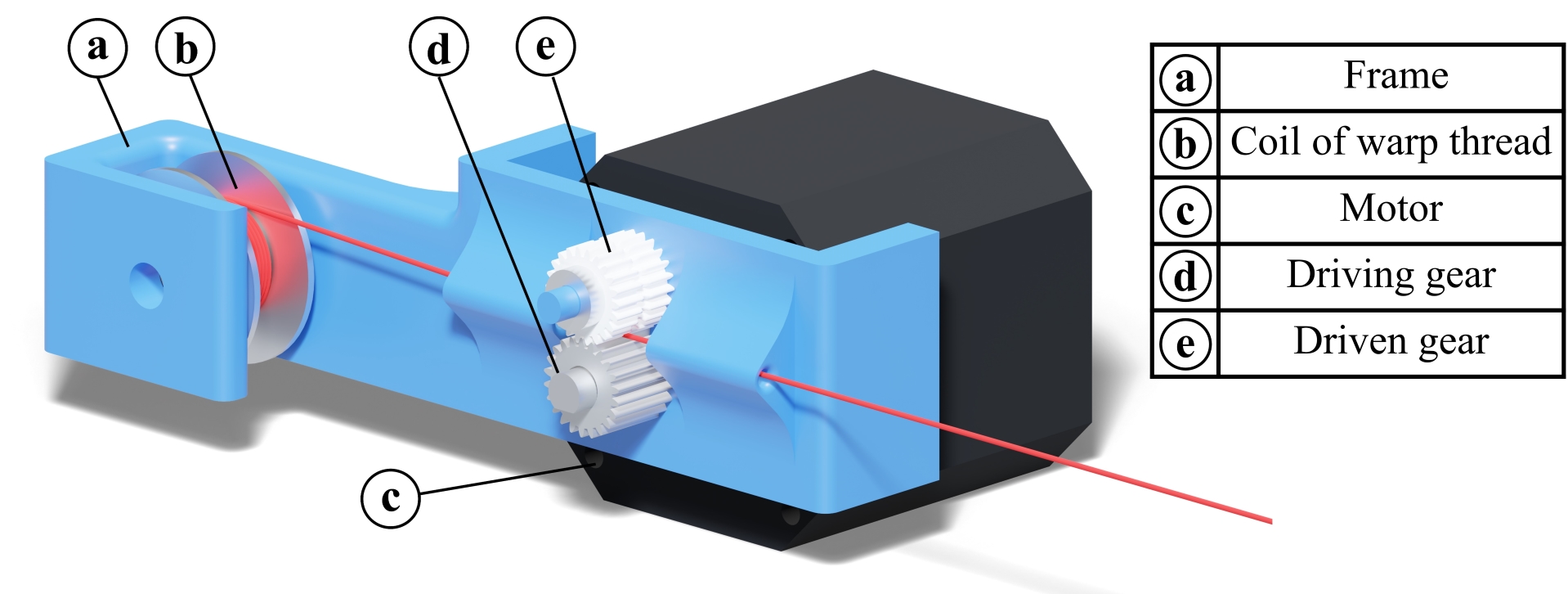}\vspace{-5pt}
\caption{Each warp beam is compounded of a frame \textcircled{a}, a coil of warp thread \textcircled{b}, a motor \textcircled{c}, and a pair of driving gears \textcircled{d} and \textcircled{e}.
}\label{fig:WarpBeam_revise}
\end{figure}

\subsection{Warp beams \rev{}{for thread length control}}\label{subsecWarpBeams}
Different from the flat weaving machines, all warp threads in our system can be individually pulled and released by the warp beams (see a matrix of warp beams shown at the right of \rev{Fig.\ref{fig:overview})}{Fig.\ref{fig:overview_machine}}. As a result, the length of each individual warp thread can be precisely controlled according to the stitch height $s_h$ to form the desired 3D freeform shape of fabrics. 

The mechanical design of a warp beam is as shown in \rev{Fig.\ref{fig:WarpBeam}}{Fig.\ref{fig:WarpBeam_revise}}. A warp thread is held by a pair of driving gears and can be released from the coil through the two spacing holes. The releasing and pulling motions of the warp thread are precisely controlled by the engaged step motor with the help of the driving gears. To control 100 warp threads, a matrix of 100 warp beams is used in our prototype machine. This is a scalable design that can be extended to many more warp beams when needed.

\subsection{Weaving mechanism}\label{subsecWeavingMech}

\begin{figure}[t]
\centering 
\includegraphics[width=\linewidth]{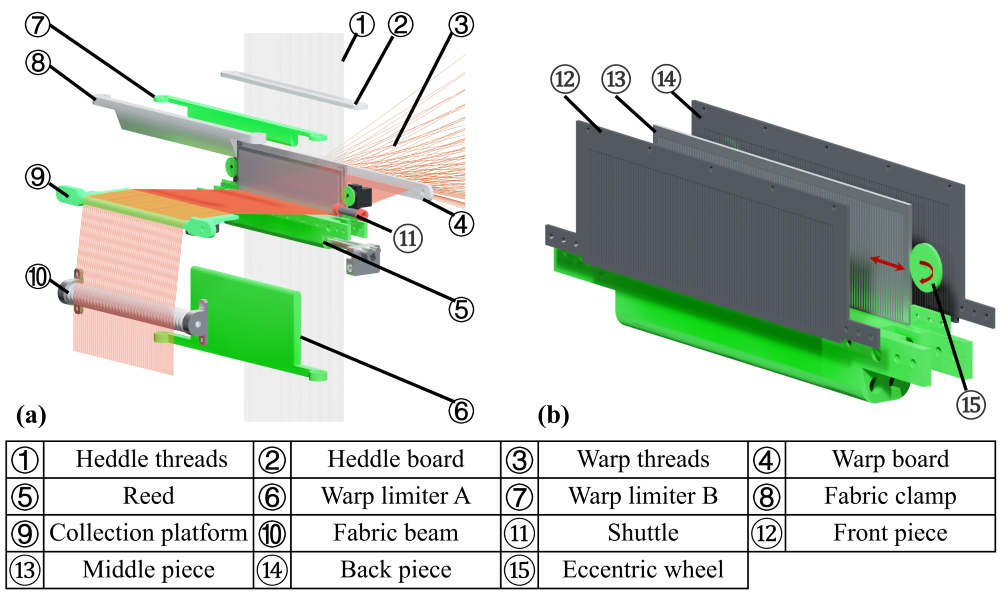}\vspace{-10pt}
\caption{A novel weaving mechanism (a) is invented in our system to form required \rev{deformation}{`push' of weaved stitches} on the fabric by operating 15 components together.
Specifically, the front piece \textcircled{\tiny{12}} and the back piece \textcircled{\tiny{14}} of the reed are fixed while the middle piece \textcircled{\tiny{13}} can shift a bit along the weft direction driven by a pair of eccentric wheels \textcircled{\tiny{15}}. The detailed structure of the reed is given in (b).
}\label{fig:weavingMechanism_revise}
\end{figure}

As the most important part of our hardware system, the weaving mechanism presented in this sub-section involves more parts and complicated motions. After forming the shed by the jacquard device and releasing the corresponding threads by the warp beams, the weaving machine completes the rest operations to form the required woven structures on the fabric. 

\rev{Different from the traditional flat machine, we design a different sequence of motions for 3D weaving, by which the newly formed woven stitches are pushed to the previous rows along the columns where the woven structures are skipped.}{During the weaving process, the newly formed woven stitches are pushed to the previous rows. With the strategy of short-row shaping, the desired 3D surface can be realized effectively.} Note that the woven structures are formed partially on each row for realizing the 3D surface shape. This pushing function for woven stitches is jointly realized by the new reed design with three pieces and the weaving motions for shifting the different lengths of threads, where the warp threads in different lengths are released by the warp beams according to the designed weaving map (from the beam side to the fabric side). The components of our weaving mechanism and the detailed structure of the reed with three pieces of blades are as shown in \rev{Fig.\ref{fig:weavingMechanism}}{Fig.\ref{fig:weavingMechanism_revise}}. The other important components include the limiters and the fabric clamp. The two limiters work like a `sliding door' that moves up and down. When the `door' is closed, all warp threads are fixed at the same level. Located above the collection platform, the fabric clamp can move down to press and hold the already completed woven structures on the fabric.

\begin{figure}[!ht]
\centering 
\includegraphics[width=\linewidth]{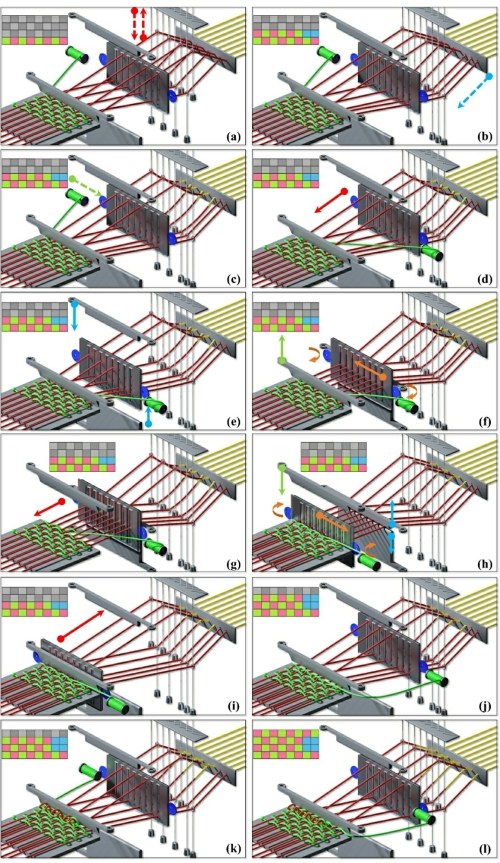}\vspace{-10pt}
\caption{The change of the woven fabric and the configurations of the weaving mechanism while fabricating a few rows as indicated by the weaving map (i.e., the colorful grids -- the meaning of different color can be found in Sec.~\ref{subsecWCode}): (a) the status after weaving the first row, (b-i) the parsing motions for weaving the second row, (j, k) the states when and after weaving the third row, and (l) the state after weaving the fourth row.
}\label{fig:weavingMotion}
\end{figure}

The parsing motions for weaving a row of woven stitches are explained below with the help of Fig.\ref{fig:weavingMotion}. 
\begin{itemize}
\item After weaving one row (see Fig.\ref{fig:weavingMotion}(a)), we are about to lift / drop the heddle threads to realize a new jacquard pattern of warp threads according to the second row of the weaving map. The changed up / down states of warp threads are as shown in Fig.\ref{fig:weavingMotion}(b). At the same time, we release the warp threads by the warp beams according to the pattern shown in the second row of the weaving map.

\item The reed is placed to a location near the heddle board while the limiters are open and the fabric clamp is pressing the fabric on the collection platform. This forms a shed with the largest area to ease the travel of the shuttle that carries the weft thread moving from one side to the other side of the loom (see Fig.\ref{fig:weavingMotion}(c, d)).

\item After the shuttle's travel, woven structures of the second row in the weaving map have already been formed on the fabric but in a loose way. We move the reed to the middle position that is slightly ahead of the limiters (see Fig.\ref{fig:weavingMotion}(d, e) -- indicated by the red arrow). 

\item The limiters are then closed to ensure all warp threads are fixed at the same height before catching them by the reed (see Fig.\ref{fig:weavingMotion}(e, f) -- indicated by the blue arrows). Meanwhile, the loosely formed woven structures are `pushed' onto the previously produced fabric by (i) shifting the middle blade of the reed to hold all warp threads (Fig.\ref{fig:weavingMotion}(f)), (ii) lifting up the fabric clamp(Fig.\ref{fig:weavingMotion}(f, g)), and (iii) tightening the loosely formed woven structures by pushing the reed forward to the position below the fabric clamp (Fig.\ref{fig:weavingMotion}(g, h)).

\item Lastly, as shown in Fig.\ref{fig:weavingMotion}(h), the fabric clamp is dropped to hold the newly formed stitches on the fabric (shown as the green arrow), the reed releases the warp threads (indicated by the orange arrows), and the limiters are opened (see the blue arrows). The reed is moved back to the initial position to prepare for weaving the next row (Fig.\ref{fig:weavingMotion}(i, j)).

%
%
\end{itemize}
These motions are repeatedly applied for every row to complete the weaving process -- e.g., see Fig.\ref{fig:weavingMotion}(j, k) for the third row and Fig.\ref{fig:weavingMotion}(l) for the fourth row. As a result, the woven fabric can be formed according to the designed 3D shape after weaving all rows.

\subsection{\rev{}{Weft length control with the help of robotic manipulation}}\label{subsecWeftLenControl}
\rev{}{After using the warp beams and the weaving mechanism to control the lengths of warp threads, we now start to tackle the other challenge of weft length control for 3D weaving. When the weft threads are longer than the length determined by the number of demanded stitches, loose loops will be formed at the end of a row. If the weft threads are too short, it will lead to unwanted distortion and lateral tension on the warp threads (also the formed stitches) at the end of a row.}

\begin{figure}[t]
\centering 
\includegraphics[width=\linewidth]{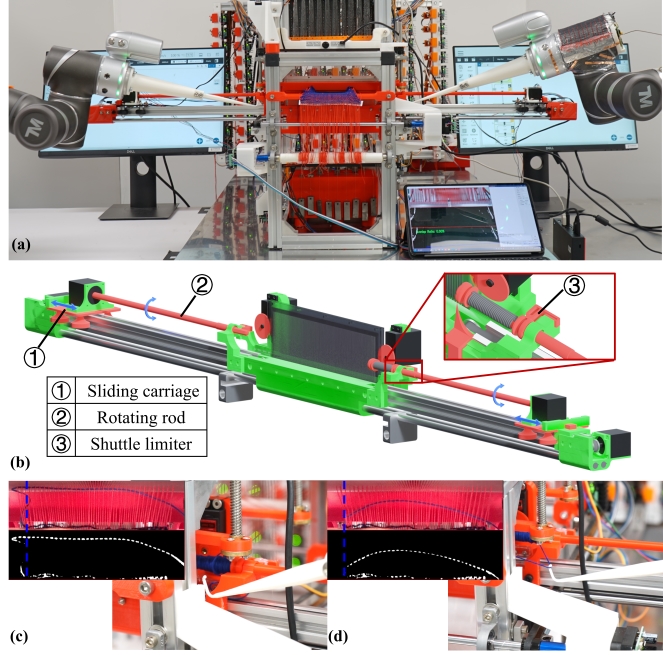}\vspace{-10pt}
\caption{\rev{}{Our system employs two robotic arms (a) to realize the length control of weft thread by working together with the shuttle moving mechanism (b). The loosely formed loop of weft thread at the end of a row will be tightened by the hook driven by a robotic arm -- see (c) and (d) for the weft thread before and after tightening. The tightening operation is driven by a camera mounted on the top of fabric clamp. The camera captured images and their results of edge extraction represented by binary images -- as vision-feedback of robotic manipulation are shown in the window views at the upper-left corner of (c) and (d), where the blue dash line indicates the target turning position of weft thread after tightening.}
}\label{fig:weftControl}
\end{figure}

\rev{}{First of all, a shuttle moving mechanism (see Fig.\ref{fig:weftControl}(b)) is designed for moving the shuttle of weft thread automatically. Two linear guide rails are installed on the left and right sides of the reed. Two rods driven by stepper motors are mounted on these guide rails to push the shuttle from left to right (by the left rod) or from right to left (by the right rod). The heads of the rods are designed in a hexagonal shape so that can engage with the shuttle and drive its rotation. While pushing the shuttle from one side to the other side by a rod, the motor of the rod will drive to rotate the shuttle so that the length of weft thread is changed. Shuttle limiters are designed to capture the shuttle at both ends, which enables the return of a rod.} 

\rev{}{Although the rotation of shuttle while traveling can release the weft thread and control its length, the resultant weft thread used for a row can be imprecise considering the diameter change of shuttle by wrapped thread. To solve this problem, we employed two robotics arms with vision-feedback control to hook onto the weft thread and pull its turning point to a demanded position as illustrated in Fig.\ref{fig:weftControl}(c) and (d).} 

In our implementation, we mount a camera on the top of fabric clamp to capture the position of the weft thread using the edge extraction based algorithm -- see the window views at the upper-left corner of Fig.\ref{fig:weftControl}(c) and (d). 
Two TM900 robotic arms are used to drive the slender hooks as their end effector, where the camera mounted on a robotic arm can help to successfully capture the weft thread by the hook. \revnew{}{The robotic arms are placed in positions by checking the reachability map to ensure that the entire motion process avoids encountering the singularity poses.} This robot-assisted tightening operation can help form stitches with better dimension control on the resultant fabrics. \revnew{}{Note that using two robotic arms here is primarily for the sake of flexibility in implementation. The similar function can be realized by using either a single robotic arm or a well-designed sophisticated mechanism, which however could be restricted by the working envelope or the possible motion flexibility.}

\subsection{Control system}

\begin{figure}[t]
\centering 
\includegraphics[width=\linewidth]{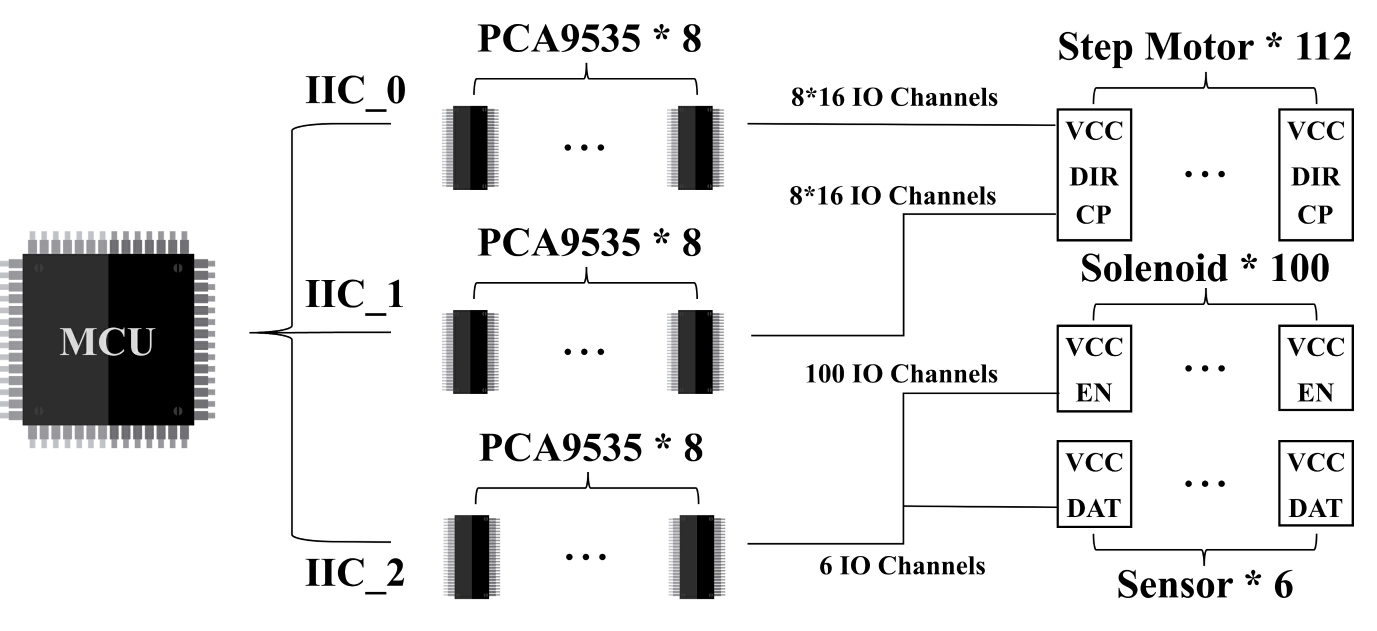}\vspace{-10pt}
\caption{A diagram of the control system used in our approach.}\label{fig:ControlSystem}
\end{figure}

The prototype of our 3D surface weaving system consists of 112 step motors (100 for warp beams, 1 for the jacquard device, and 11 for the weaving mechanism), 100 electronic solenoids (for the warp threads), and 6 position sensors. It is important to control these actuators and sensors in a stable and synchronized way. As over 300 IO channels are required, the IO extension chip (i.e., PCA9535) is employed to connect the sensors / actuators to the \textit{Micro Control Unit} (MCU -- i.e., F103ZET6) through the \textit{Inter-Integrated Circuit} (IIC). The diagram is as shown in Fig.\ref{fig:ControlSystem}, where we use three sets of IIC to communicate with 24 {PCA9535} to complete the control tasks -- considering that the maximal number of chips for each IIC is 8. The firmware of our control system was developed in C programming language and all circuits are designed by using EasyEDA. Detailed schematics and PCB layouts have been provided in Appendix A.

\begin{figure*}[t]
\centering 
\includegraphics[width=\linewidth]{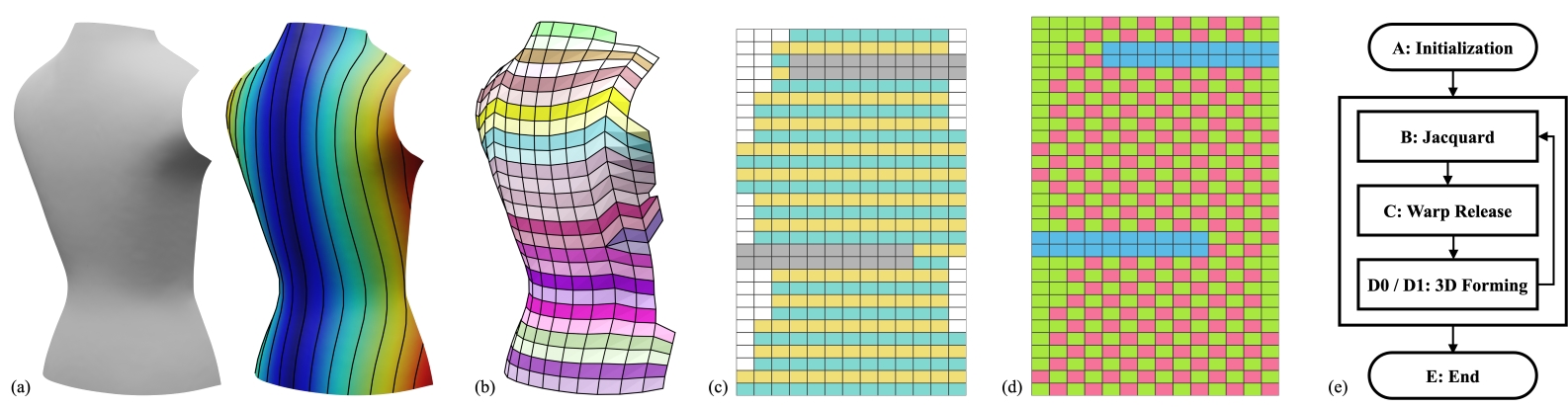}\vspace{-8pt}
\caption{Given a target surface, we first generate a geodesic distance-field by using the central line of the piece as source (a). A stitch mesh for short-row shaping based knitting as shown in (b) can be generated from the isocurves of the geodesic distance-field by optimized quadrangles / triangles (details can be found in \cite{liu2021knitting}). The stitch mesh can directly map into (c) a knitting map $K(\cdot, \cdot)$, which is converted into (d) a weaving map $W(\cdot, \cdot)$ and the machine code with its structure / syntax as explained in (e) to supervise the operations of our weaving machine.
}\label{fig:software}
\end{figure*}

\section{Machine Code Generation}\label{secSoftware}
Given a 3D freeform surface to be fabricated by our weaving machine, we first generate a short-row based stitch mesh that can result in a knitting map. After that, the knitting map is converted into machine code including both the information of jacquard pattern and the instructions for weaving operations. The whole pipeline of machine code generation has been illustrated in Fig.\ref{fig:software}.

\subsection{Knitting map generation}\label{subsecKnittingMap}
A geodesic distance-field is first computed from the user-defined source (point or curve) on the input surface, where the isocurves of the field with the distance as the target stitch width $s_w$ are extracted to serve as the column boundaries of stitches (see Fig.\ref{fig:software}(a) for an example). The isocurves are then sampled into segments with lengths equal to the target stitch height $s_h$. 
Lastly, the stitch meshes are constructed by connecting segments on neighboring isocurves with optimized quadrangles / triangles. The following rules are applied to generate the knitting maps considering manufacturing constraints:
\begin{enumerate}
\item The stitches are organized row by row;
\item All stitches in the same row are neighboring connected;
\item Stitches in the same row are formed in an alternating left-to-right and right-to-left order;
\item The ending stitch of the current row should be neighboring to the starting stitch of the next row.
\end{enumerate}
The stitch mesh generated by the algorithm in \cite{liu2021knitting} can satisfy all these requirements. As a result, every stitch can be mapped to a location in the 2D knitting map by its row index and its location in the row -- see Fig.\ref{fig:software}(b, c) for an example.

A knitting map presents the operational information about how to form stitches while traveling in the alternated left-to-right and right-to-left ways by the carriers holding yarns. Three different blocks displayed in different colors  are defined in a knitting map (see Fig.\ref{fig:software}(c)). The cyan / yellow blocks are the stitches formed by loops of yarns in knitting, where the cyan ones are stitches formed by the carrier traveling from right to left and the yellow blocks represent those formed by traveling from left to right. No stitch will be formed for the white or the gray blocks. That means the carrier will \textit{not} travel into the white / gray regions. Due to the short-row shaping technique, the stitches below and above the gray blocks will be connected to locally change the length of fabrics. As a result, 3D surfaces are produced by the connected stitches.

\subsection{Converting knitting map into weaving code}\label{subsecWCode}
Different from knitting, every stitch of the woven structure is formed together by two neighboring warp threads and two neighboring weft threads. Therefore, for a knitting map $K(\cdot)$ with $M$ columns and $N$ rows, it will be converted into a weaving map $W$ with $(M+1)$ columns and $(N+1)$ rows. When fabricating the $i$-th row of the woven structure, $W_{i,j}$ provides information for both the Jacquard configuration of the $j$-th warp thread and the corresponding warp beam's operation. 

Unlike the short-row shaping of knitting that the carrier's travel is only conducted in a limited range for each row (e.g., only the region with yellow or cyan blocks in Fig.\ref{fig:software}(c)), the carrier holding the weft thread always travels from the head to the tail of each row on the loom. When two neighboring warp threads have the same Jacquard configuration (i.e., both \textit{up} or both \textit{down} as $W_{i,j}=W_{i,j+1}$), no woven structure can be formed between them. On the other aspect, two different results can be obtained by using different operations of the warp beam for the $j$-th thread when $W_{i,j}=W_{i+1,j}$:
\begin{itemize}
\item If the beam releases the $j$-th thread with a unit length $s_h$ (i.e., woven stitch height) as usual, it will become a floating thread between the $i$-th row and $(i+1)$-th row;

\item If the beam keeps holding the thread, the already formed woven stitch before the $i$-th row will be tightly `sewn' to the later stitch after the $(i+1)$-th row.
\end{itemize}
The first situation happens for the white regions in the knitting map, and the second case occurs in the gray regions.

An algorithm is developed to generate $W(\cdot)$ from $K(i,j)$ with $i=1,\ldots,N$ as row index (starting from the bottom) and $j=1,\ldots,M$ as column index (starting from the left). For the sake of explanation, colors are used while introducing the algorithm below (with Fig.\ref{fig:software}(c, d) as an example).
\begin{itemize}
\item \textbf{Step 1:} Create a new weaving map as $W(i,j)$ with $(N+1)$ rows and $(M+1)$ columns and fill all blocks of $W(i,j)$ by green. 

\item \textbf{Step 2:} We check the color of every block $K(i,j)$ in the knitting map so that 1) $W(i+1,j)$ is assigned as magenta when $K(i,j)$ is yellow or cyan or 2) $W(i+1,j)$ is assigned as blue when $K(i,j)$ is gray.

\item \textbf{Step 3:} For every row $W(i,\cdot)$ ($i \neq 1$), if the last non-green block is located at $W(i,j')$, we assign $W(i,j'+1) = W(i,j')$ -- i.e., one column is extended. 

\item \textbf{Step 4:} For the first row $W(1,\cdot)$, we copy it from the second row as $W(1,j)=W(2,j)$ for every blocks.

\item \textbf{Step 5:} For a magenta block $W(i,j)$, we convert its color to green when 1) both $i$ and $j$ are odd numbers or 2) both $i$ and $j$ are even numbers.
\end{itemize}
After running this algorithm, we can generate a weaving map with information stored for both the Jacquard pattern and the warp beams' operations. Specifically, we have
\begin{enumerate}
\item $W(i,j)=\mathrm{blue}$: the $j$-th warp beam will not release the warp thread and the Jacquard pattern of the $j$-th warp thread is kept as `down' state;
\item $W(i,j)=\mathrm{magenta}$: the $j$-th warp beam will release the thread for a  length $s_h$ as woven stitch height and the Jacquard pattern of the $j$-th thread is changed to `up' state if it is not;
\item $W(i,j)=\mathrm{green}$: the $j$-th warp beam will release the thread and the $j$-th thread is dropped to the `down' state for its Jacquard pattern.
\end{enumerate}
An example weaving map can be found in Fig.\ref{fig:software}(d).

Based on the above definitions for different colors, we convert the weaving map into a \textit{W-code} file with a structure as shown in Fig.\ref{fig:software}(e) to supervise the operations of our 3D surface weaving machine. The syntax of \textit{W-code} consists of five commands:
\begin{itemize}
\item \textbf{A:} The initialization command to reset the hardware for the coming weaving process.
\item \textbf{B:} The command is given together with $(M+1)$ binary numbers to instruct the Jacquard device to selectively lift the corresponding warp threads (e.g., Fig.\ref{fig:weavingMotion}(b)).
\item \textbf{C:} A command followed by $(M+1)$ binary numbers is used to instruct the warp beams for selectively releasing the corresponding warp threads, where the length of release is defined according to the target woven stitch's height $s_h$.
\item \textbf{D:} A command followed by a boolean value for driving the shuttle (i.e., D0 -- from left to right and D1 -- from right to left). The command completes the operations of weaving one row as described in Sec.~\ref{subsecWeavingMech} (see also Fig.\ref{fig:weavingMotion}(c)-(i)).
\item \textbf{E:} The end command to stop the weaving process.
\end{itemize}
\rev{}{Note that this W-code in general can be employed on weaving machines in different configurations with carefully designed controllers.} An example W-code file for the weaving map shown in Fig.\ref{fig:software}(d) can be found in Appendix B.

\begin{figure*}[t]
\centering 
\includegraphics[width= \linewidth]{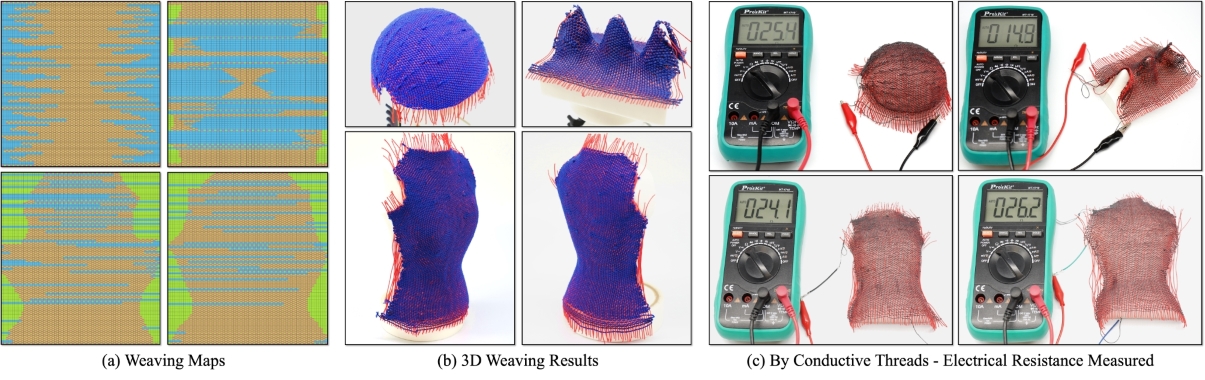}\vspace{-10pt}
\caption{We tested our 3D surface weaving machine on four different examples, where their weaving maps are given in (a). The results of physical fabrication are shown in (b). Our weaving system enables the flow of electric current over the freeform surfaces of fabrics by using conductive threads in (c).
\revnew{}{The measurements obtained with the avometer represent the resistance between two ends of the weft thread, which indicates the continuity of the conductive weft thread.}}\label{fig:FabResultsAll}
\end{figure*}

\begin{table}[t]
\caption{Statistics of Computation}\vspace{0pt}\centering\label{tab:CompStatistics}\footnotesize
\begin{tabular}{r|r||c|c|c|c|c}
\hline
        &  Trgl.  &   &  \multicolumn{2}{c|}{Thread Num.$^{\dag}$} & \multicolumn{2}{c}{Comp. Time$^{\ddag}$ (sec.)} \\
\cline{4-7}
Model  &  Num. &  Fig. & Weft & Warp & Knit Map & W-Code \\
\hline \hline
Hemisphere   & $4000$  & \ref{fig:overview_machine} &  $138$ & $72$ & $1.64$ & $0.38$\\
            &   & - &  $276$ & $144$ & $1.67$ & $1.71$\\
\hline
Vest-Front &  $26593$ & \ref{fig:teaser}  &  $170$ & $71$ & $1.78$ & $0.49$\\
    &   & - &  $340$ & $142$ & $7.01$ & $2.01$\\
\hline
Vest-Back  & $28763$  & \ref{fig:FabResultsAll}  &  $173$ & $72$ & $1.97$  & $0.52$\\
   &   & - &  $346$ & $144$ & $7.75$  & $2.19$\\
\hline
Triple-Peak  & $10016$  & \ref{fig:FabResultsAll}  &  $190$ & $58$ & $0.97$  & $0.45$\\
   &   & - &  $380$ & $116$ & $3.85$  & $1.98$\\
\hline
\end{tabular}
\vspace{-5pt}
\begin{flushleft}
$^\dag$~Tests with different number of threads were conducted to demonstrate the scalability of our computational method although the maximally allowed number of warp threads on our prototype machine is only 100.\\
$^\ddag$~The computing time of W-code generation includes both the step of weaving map generation and the step for converting it into W-code.
\end{flushleft}
\end{table}

\section{Results and Discussion}\label{secResult}
We have made a prototype machine by using off-the-shelf electrical / mechanical components, motors, and 3D printed parts. The software of our system was developed by Python. The experimental tests reported in this section were all run on a desktop PC with AMD Ryzen 7 5800X 3.8GHz CPU and 32GB RAM. Different materials are employed to fabricate examples with a variety of freeform 3D shapes. A supplementary video to demonstrate the function of our work can be accessed at: \url{https://youtu.be/mzdlq5Sk8NQ}.

\subsection{Statistics of computation and fabrication}
We have tested our prototype system in physical experiments by fabricating models as shown in Fig.\ref{fig:FabResultsAll}. The testing results are quite encouraging although both the \rev{hardware}{machine} and the \rev{software}{weaving-map generation algorithm} can be further optimized. \rev{}{The limitations of our current implementation are discussed in Sec.~\ref{subsecLimitations}}

The computational statistics of these examples have been listed in Table \ref{tab:CompStatistics}, where the examples are tested by using threads in two different resolutions. As it can be observed, the computation on all tests can be completed within 10 seconds and the major bottleneck is the step of knitting map generation \cite{liu2021knitting}. The complexity of the knitting map generation algorithm is $O(AB\log B)$  with $A$ and $B$ being the maximal number of stitches along the row and the column directions on the knitting map respectively.

\begin{figure}[t]
\centering 
\includegraphics[width=\linewidth]{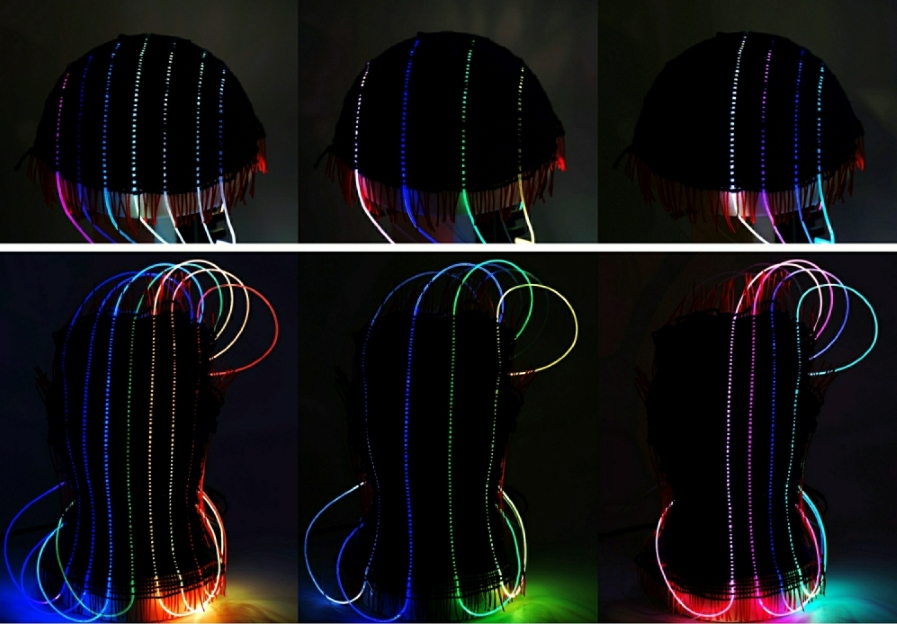}\vspace{-5pt}
\caption{Examples to demonstrate the functionality of 3D surface weaving with optical fibres.
}\label{fig:ResOptFab}
\end{figure}

We have physically fabricated these example models on our prototype machine by using different materials. While the waxed threads with cotton core are employed for warp threads, both the cotton thread (waxed) and the conductive threads (bare copper with silicone rubber insulation) have been used for weft threads in our tests (see Fig.\ref{fig:FabResultsAll}). The fabrication of all models can be completed within a reasonable time as listed in Table \ref{tab:FabStatistics}. We have tested the examples by replacing a few warp threads with optical fibres -- named as `hybrid' warp threads in Table \ref{tab:FabStatistics} with the results shown in Figs.~\ref{fig:overview_machine}~\&~\ref{fig:ResOptFab}. \rev{}{Our weaving process and machine do not have special requirements on the materials to be used as threads for forming woven structures as 3D surfaces. A material can be used as long as it can be successfully extruded by the warp beams and winded by the shuttle.}

\begin{figure}[t]
\centering 
\includegraphics[width=\linewidth]{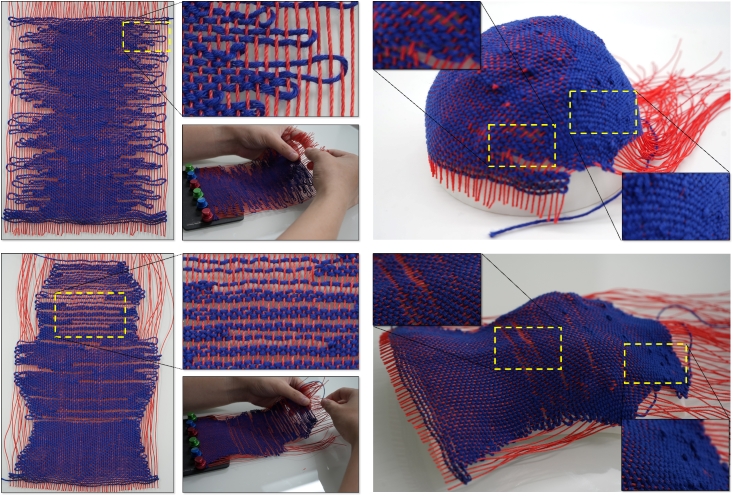}\vspace{-5pt}
\caption{Examples to demonstrate the problems caused by \rev{handmade}{manual} deformation on the fabrics made by flat weaving, where the height of woven stitches $s_h$ cannot be well controlled. The heights are non-uniformly distributed -- i.e., larger at the fixed end while shorter at the pushing side.
}\label{fig:ResHandmade}
\end{figure}

\begin{figure*}[t]
\centering 
\includegraphics[width=\linewidth]{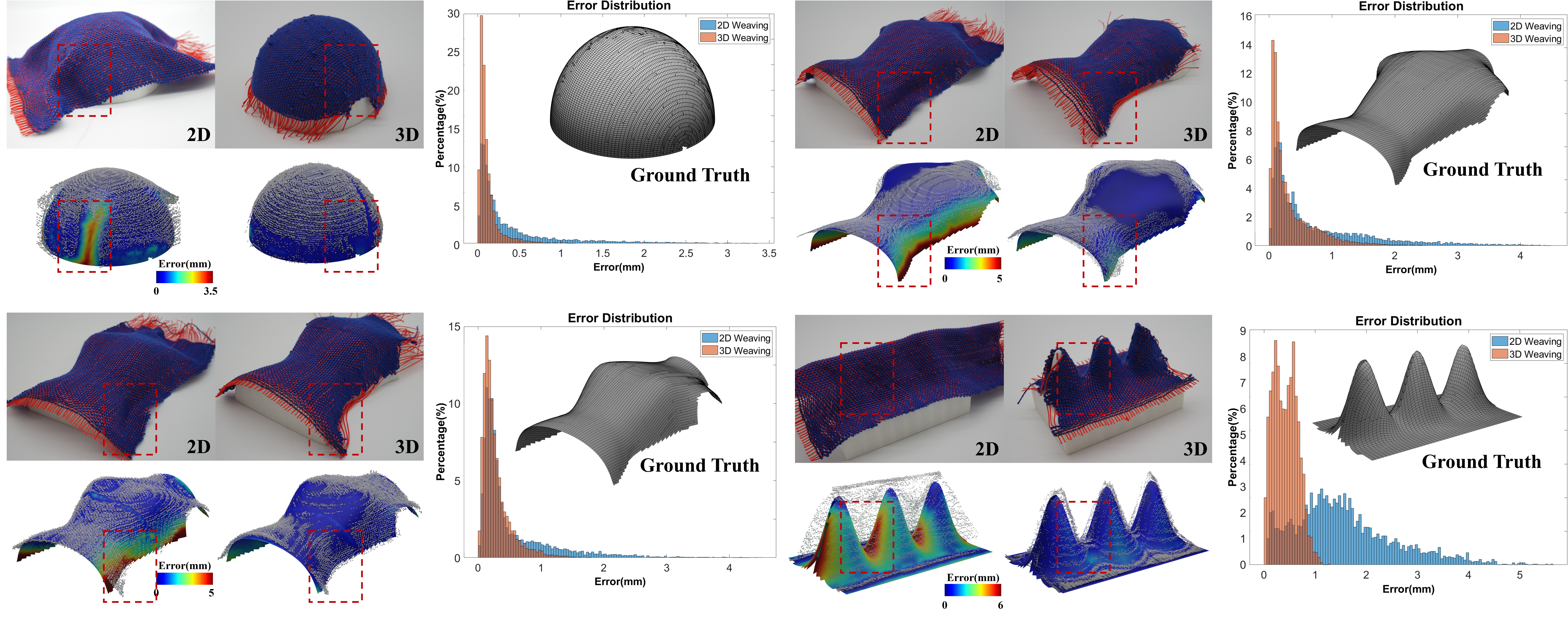}\vspace{-5pt}
\caption{Comparison of shape approximation error by the distribution of distances between sample points on the given surfaces to the 3D point clouds scanned from the fitting results of flat woven fabric (2D) and those by our surface weaving (3D), where the errors are displayed by the color maps and the error histograms. From top to bottom, the models are hemisphere (6,575), vest-front (8,220), vest-back (7,989) and triple-peak (4,057), where the numbers of samples for each model are given in the bracket. \rev{}{Regions with large error on the 2D results and their corresponding regions on the 3D results are highlighted by rectangles drawn in red dash lines.}
}\label{fig:ResGeo}
\end{figure*}

\begin{table}[t]
\caption{Statistics of Physical Fabrication}\vspace{0pt}\centering\label{tab:FabStatistics}\footnotesize
\begin{tabular}{r||c|c||c|c||c}
\hline

 & \multicolumn{2}{c||}{Thread \#} & \multicolumn{2}{c||}{Thread Materials} & Weaving \\
\cline{2-5}
Model  &  Weft & Warp & Weft & \multicolumn{1}{c||}{Warp} & Time (min.) \\
\hline \hline
   &  &  & Cotton &  Cotton & $140$ \\
Hemisphere           &  $138$ & $72$ & Conductive  & Cotton & $152$ \\
   &  &  & Cotton &  Hybrid & $145$ \\
\hline
   &  &  & Cotton & Cotton & $170$ \\
Vest-Front           &  $170$ & $71$ & Conductive  & Cotton & $181$ \\
 &   &  &  Cotton &  Hybrid & $173$ \\
\hline
   &  &  & Cotton & Cotton & $174$ \\
Vest-Back            &  $173$ & $72$ & Conductive   & Cotton & $183$ \\
\hline
   &  &  & Cotton & Cotton & $195$ \\
Triple-Peak           &  $190$ & $58$ & Conductive  & Cotton & $211$ \\
\hline
\end{tabular}
\end{table}

\subsection{Automatic vs. manual \rev{}{deformation based} shape forming}
One of the major contributions of our hardware design is the warp beams and the weaving mechanism that can precisely control the lengths of warp threads therefore also the height of woven stitch $s_h$. This is also the reason why our machine can form 3D surfaces automatically. We demonstrate this automatic 3D forming function by comparing with examples that are fabricated by conventional flat weaving. 

The weaving map can still be implemented on a commercial flat weaving machine with a jacquard device. Gaps (actually floating threads) will be formed on the fabrics between rows (see the zoom view in the left of Fig.\ref{fig:ResHandmade}). Although we can manually pull the end of the warp threads to eliminate these gaps and deform the fabric, these operations cannot accurately control the height distribution of woven stitches along the pulled warp threads. For manual 3D shape forming, we fix one end of the flat woven fabrics while pulling warp threads and pushing weft thread on the other side (see also Fig.\ref{fig:ResHandmade}). Note that the pushing force is applied everywhere along the same warp thread and the gaps are eliminated by squeezing neighboring stitches. The result of manual form generally shows larger stitch height at the fixed end but smaller stitch height near the pushing side. As a consequence, the target 3D shape of woven fabrics cannot be accurately formed. 

Additionally, for the surface boundary that are next to the large green regions in the weaving map (\rev{}{e.g.,, the bottom left one shown in Fig.\ref{fig:FabResultsAll}(a)}), the manual operation will lead to large distortions -- see the waist region for the vest-front model \rev{}{in Fig.\ref{fig:ResHandmade}}. In short, the manual 3D shape forming cannot well control the boundary shape of fabrics.

\subsection{Analysis of geometric errors: 3D shape}
One of the most important purposes of this work is to reach a high level of geometric accuracy on the 3D surface shape of the produced woven fabrics. In the conventional textile industry, a flattening algorithm (e.g., \cite{TASE2011,CAD2005}) was employed to compute a (trimmed) 2D panel shape to fit onto the desired 3D object (e.g., human bodies in the apparel industry). However, as 3D surfaces are in general not \textit{flattenable} (or named as \textit{developable} in geometric modeling), wrinkles will appear at regions where the curvature is large and leftovers are inevitable. For a better fit, smaller pieces of the flat-woven fabric are generated and stitched together, which involved time-consuming sewing processes that are labor-intensive (e.g., \cite{zhao2023evolutionary, zhang2019FabFormwork, TVCG2008}). 

We conducted experimental tests to study the shape approximation errors on a variety of models. Both the skinned shape by flat woven fabrics and the shape from 3D surface weaving are compared with the target shape (as the ground truth). A structured light based 3D scanner \cite{sacnner} is employed to generate the point clouds for each surface, which is also placed into a well-registered pose w.r.t. the ground truth shape. The shortest distances between the sampled points on the given surface to the scanned point clouds are employed as shape approximation errors, which are visualized as color maps and histograms in Fig.\ref{fig:ResGeo}. From the results of error analysis, we can easily conclude that models fabricated by our 3D surface weaving have much smaller errors in all examples -- i.e., mainly less than 1mm as can be observed from the histograms. When the surface to be fabricated becomes more and more complex, the geometric error on fitting results by flat woven fabrics will increase significantly (see the triple-peak model in Fig.\ref{fig:ResGeo} for an example). 

\begin{figure}[t]
\centering 
\includegraphics[width=\linewidth]{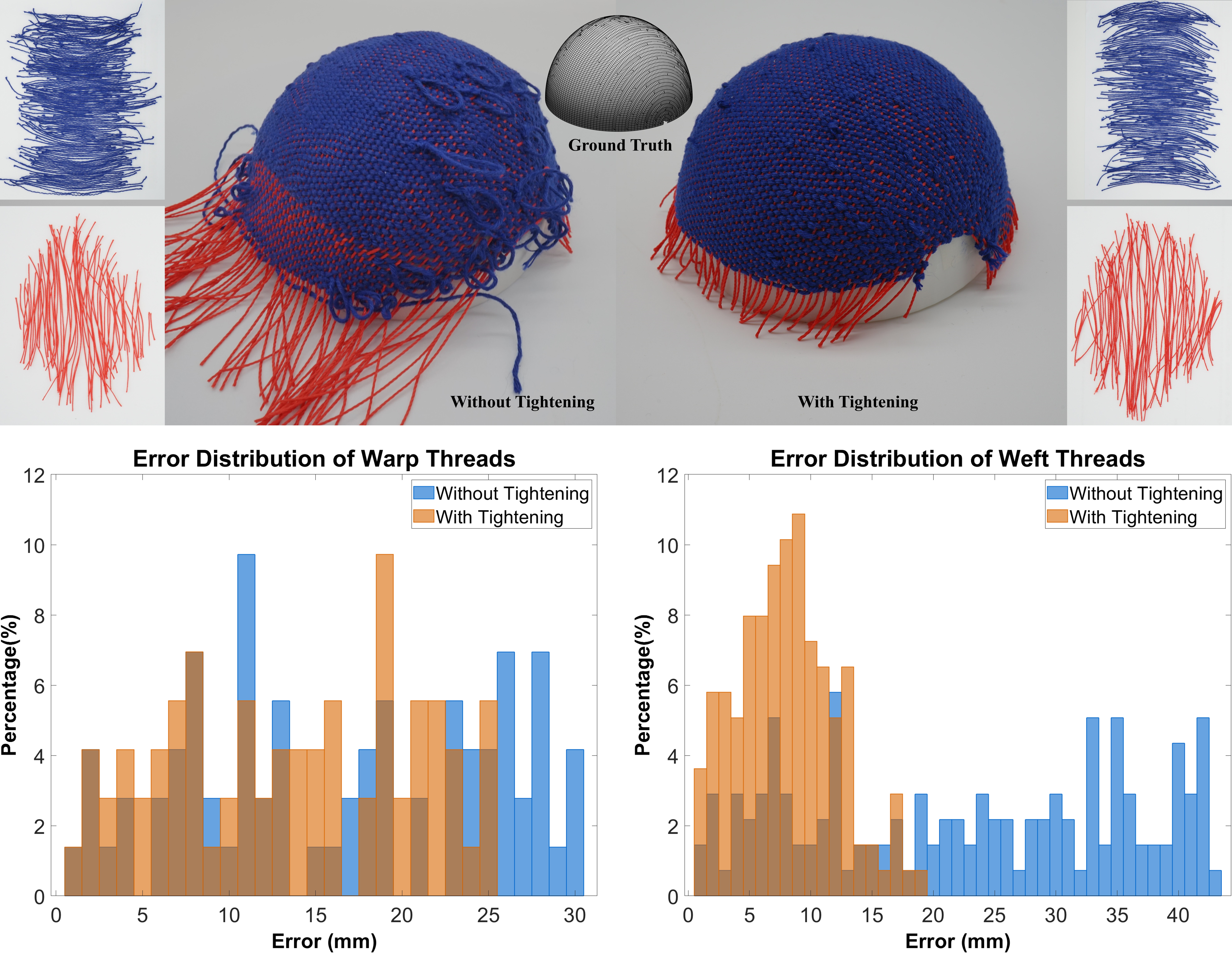}\vspace{-5pt}
\caption{\rev{}{Experimental test to study the length accuracy of 
warp / weft threads on the resultant 3D fabrics, where the lengths of every warp and weft threads are measured and compared with the demanded lengths. The error analysis as distribution has been given for both the results with vs. without the robotic tightening manipulations.} }\label{fig:ResThreadLengthCtrl}
\end{figure}

\subsection{\rev{}{Analysis of geometric errors: lengths of threads}}
\rev{}{Experiments are conducted to evaluate the performance of our method in the length control for both the warp threads and the weft threads on the resultant 3D fabrics produced by our system. We measure the lengths of every warp and weft threads and compare their lengths with the demanded lengths, which are specified by the weaving map. The tests are given for the hemisphere model as it does not have concave boundaries, considering the lengths of warp threads across concave boundary are difficult to measure accurately.}

\rev{}{Ablation study has been conducted to fabricate two hemispherical fabrics with vs. without the robotic manipulation based tightening process after weaving the stitches of each row. Specifically, the fabrics made by our 3D weaving machines are cut into threads at the end of each row and column to measure the lengths of all threads. From the error analysis shown in Fig.\ref{fig:ResThreadLengthCtrl}, we find that large variations of length errors for weft threads are observed on the fabric without the robotic tightening process. At the same time, slightly worse length control can also be observed on warp threads. This is mainly because that, without the robotic tightening, the uncontrolled weft lengths near the end of a row will also lead to distorted stitches with varied warp lengths.}

\subsection{Analysis of geometric errors: thread-paths}
Besides of accurate 3D shape, our 3D surface weaving approach has another intrinsic advantage -- the warp threads naturally preserve the geodesic distance between each other. This enables the capability to embed functional threads at equal distances when employing them as warp threads. We demonstrate this by using optical fibres in our experiment and comparing them to the results obtained from flat weaving. 

\begin{figure}[t]
\centering 
\includegraphics[width=\linewidth]{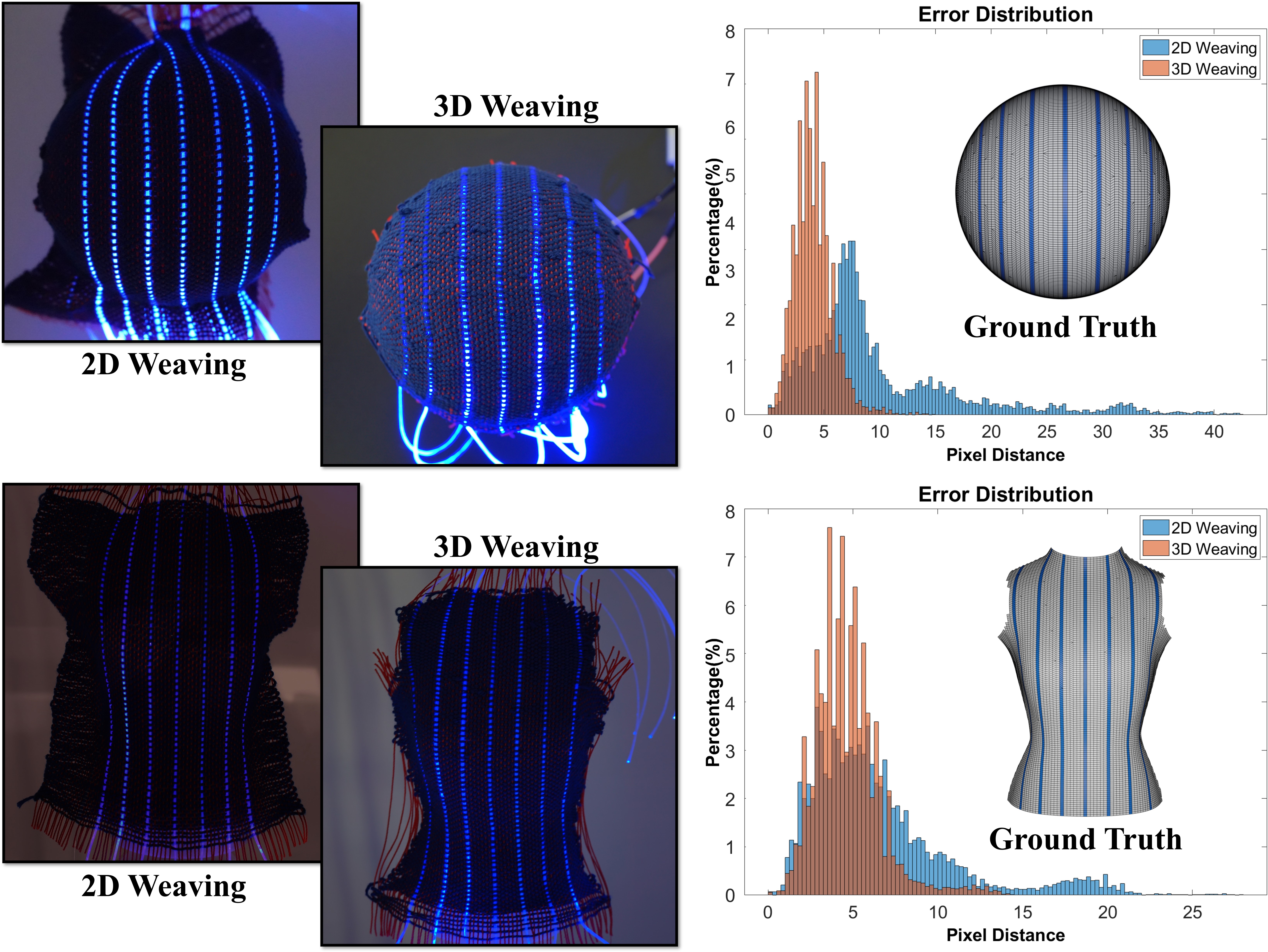}\vspace{-5pt}
\caption{Results to demonstrate the accuracy of warp thread-paths that are formed on the fabrics produced by the flat weaving (2D) vs. the surface weaving (3D). We target on distributing optical fibres on a given freeform surface at an equal distance. The statistical errors are shown in the histogram with the unit as pixels in the image space (i.e., in the resolution of $652 \times 850$ for an area of $ 150\mathrm{mm} \times 195\mathrm{mm}$).}\label{fig:ResPath}
\end{figure}

As shown in Fig.\ref{fig:ResPath}, we aim to insert optical fibres with equal distance on the hemisphere and the vest-back models. The ground-truth distribution / locations of these curves are illustrated as blue curves on the 3D models with the distance as $7 s_w$ between neighboring curves. For weaving such fabrics, we replace 7 among all warp threads with optical fibres while remaining 6 warp threads between every neighboring optical fibres. Note that the distribution of fibres is selected in a symmetric way with the middle fabric coincident to the symmetric plane of the models. The same setup of fabrication is applied to both flat weaving (2D) and surface weaving (3D). After fitting the resultant fabrics onto the target models (registered by the central-line curve), we measure the light distribution of the optical fibres by photos to evaluate the distribution of the corresponding thread-paths. 

Specifically, for each optical fibre, a B-spline curve is fit onto the lighting points extracted from photos and the curve is re-sampled into $1,000$ points. The distances between these sample points and the target ground-truth curves are measured and analyzed as shown in the histogram. The tests are conducted on results obtained from both 2D and 3D weaving. Apparently, the fitting from 2D fabrics onto 3D models generates uncontrolled distortion -- i.e., the resultant thread-paths of optical fibres present much larger errors. 

\subsection{Continuity of Threads} 
Our 3D surface weaving machine can help produce functional fabrics with many interesting applications in mechatronics. When conductive threads are used (as already demonstrated in Fig.\ref{fig:FabResultsAll}(c)), our 3D surface weaving approach enables the flow of electric current over the surfaces of fabrics so that they can work like soft PCBs. This has many potential applications such as functional garments \cite{9788480}, motion sensing \cite{9263352}, and `skin' of cobot \cite{Pang2021CoboSkin}. To enable this function, continuity on both the warp and the weft threads is required. While forming a 3D freeform shape by our surface weaving technique, we use a single thread for weft and multiple continuous threads for warp directions. Continuity is preserved on the resultant woven fabrics. 

\begin{figure}[t]
\centering 
\includegraphics[width=\linewidth]{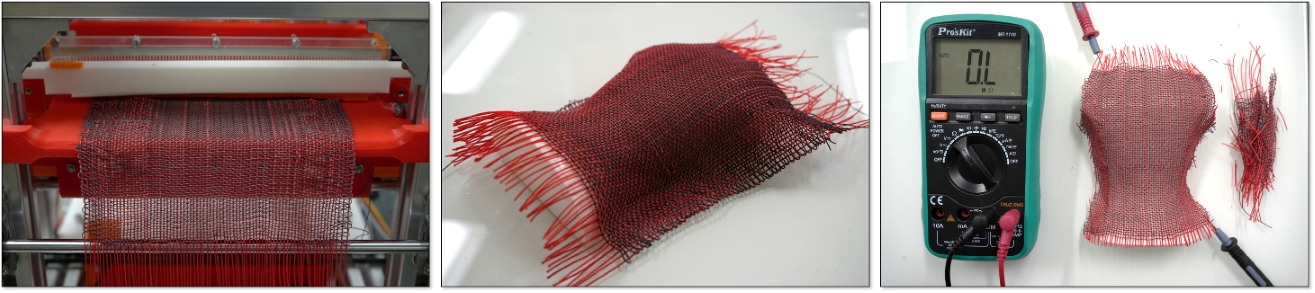}\vspace{-8pt}
\caption{Experiment to demonstrate the flat weaving technique cannot achieve both the continuity of conductive thread and the perfect fitting 3D shape together. An infinity is measured as the electrical resistance after trimming the flat fabrics to form a target 3D shape for the vest-back -- i.e., the conductive weft thread has been broken.}\label{fig:ResRes}
\end{figure}

The following experiment is to demonstrate the problem of discontinuity when flat woven fabrics are employed. First of all, a piece of woven fabric is made by flat weaving by using a conductive thread for weft directions (the left of Fig.\ref{fig:ResRes}). Because there are excess boundaries and winkle regions when fitting onto the mold (see the middle of Fig.\ref{fig:ResRes}), the 2D fabric has to be trimmed for a better fit. After fitting and trimming, most of the weft threads are cut and the circuits within the fabric are open which leads to infinity as the measured electrical resistance (the right of Fig.\ref{fig:ResRes}). Differently, our surface weaving technique can generate a perfect fitting shape while preserving the continuity of the conductive weft thread (see the back-vest example shown in Fig.\ref{fig:FabResultsAll}). 

\subsection{\rev{}{Limitations}}\label{subsecLimitations}
One major limitation of our current implementation is that two robotic arms are employed for realizing the length control of weft threads. This significantly increases the cost and the complexity of the manufacturing system. A possible future work is to \revnew{design a special mechanism equipped with force sensors to complete the tightening step for weft threads.}{utilize sword bar mechanical structures to control the length and distribution of the weft thread while also using a camera for visual feedback and control.}


\rev{}{At present, the weaving speed of our prototype machine is much slower than those flat weaving machines used in industry. We choose very conservative speeds in many steps of motions on our prototyping machine, which can be further tuned and optimized. Although 3D weaving involves steps and motions much more complicated than flat weaving, the mechatronics design of our machine can still be further optimized for improving the efficiency of 3D weaving.} 

\begin{figure}[t]
\centering 
\includegraphics[width=\linewidth]{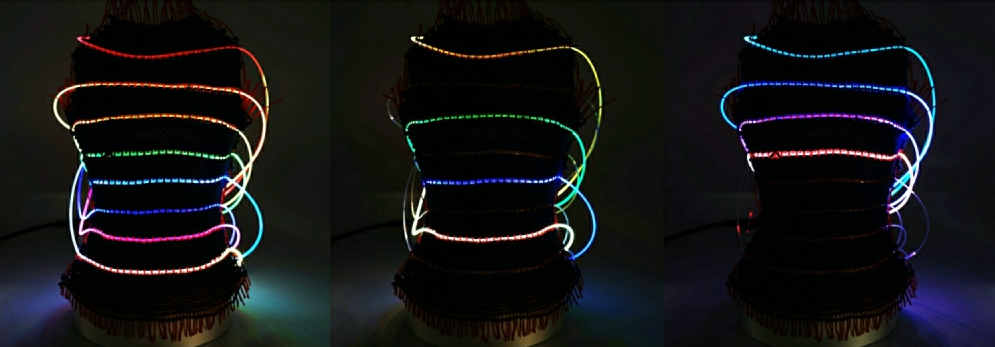}\vspace{-5pt}
\caption{An example of embedding optical fibres along the weft directions by the technique of in-lay. 
}\label{fig:ResInLay}
\end{figure}

\begin{figure}[t]
\centering 
\includegraphics[width=\linewidth]{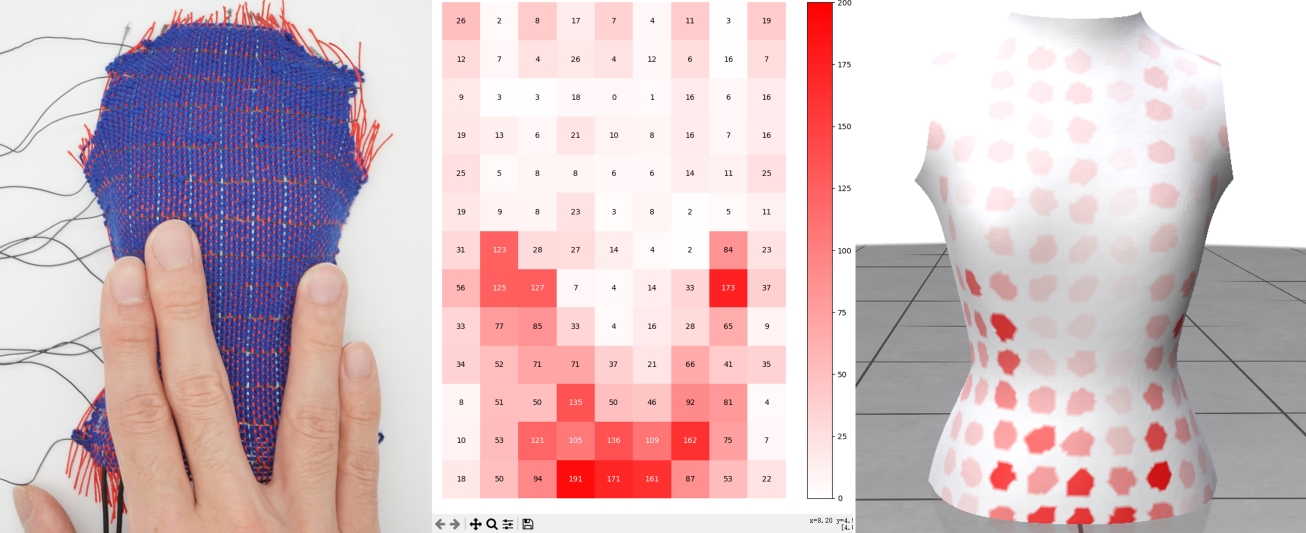}\vspace{-5pt}
\caption{\rev{}{An example of vest-front fabric made by using 1) bare stainless steel wire for warp threads and 2) silicone insulated copper wire for weft threads (left), which can be developed into a touch sensing interface by capacitance \cite{Lotters99capacitance}. (Middle) the distribution of touch that is sensed by the grid of conductive threads, and (right) the corresponding touch visualized on the 3D surface.} 
}\label{fig:resMetalWiresBothDir}
\end{figure}

In the examples listed above, the threads with high stiffness (such as optical fibres) are only employed as warp threads. This is because the weft thread needs to form U-turn at the end of each row. However, we can employ the in-lay technique to `insert' an additional thread (with high bending-stiffness) between those neighboring rows formed by a continuous weft thread. An example can be found in Fig.\ref{fig:ResInLay}. Note that the in-lay fabrication needs manual operations on our prototype machine \rev{although it has the potential to be automated by robotic techniques (e.g., \cite{Petrik2020FabFolding,She21CableMani})}{in our current experiment}. \rev{}{However, when the U-turn at the end of each row is tolerable on materials with high bending-stiffness (e.g., metal wires), our method can produce fabrics by using these stiff materials in both warp and weft threads -- see Fig.\ref{fig:resMetalWiresBothDir} for an example by using these conductive grids to form a capacitance-based touch sensing interface \cite{Lotters99capacitance}.} Lastly, the abrasion of warp threads caused by the reed's motion requires the threads to be waxed to prevent damage to its `core'.

\begin{figure}[t]
\centering 
\includegraphics[width=\linewidth]{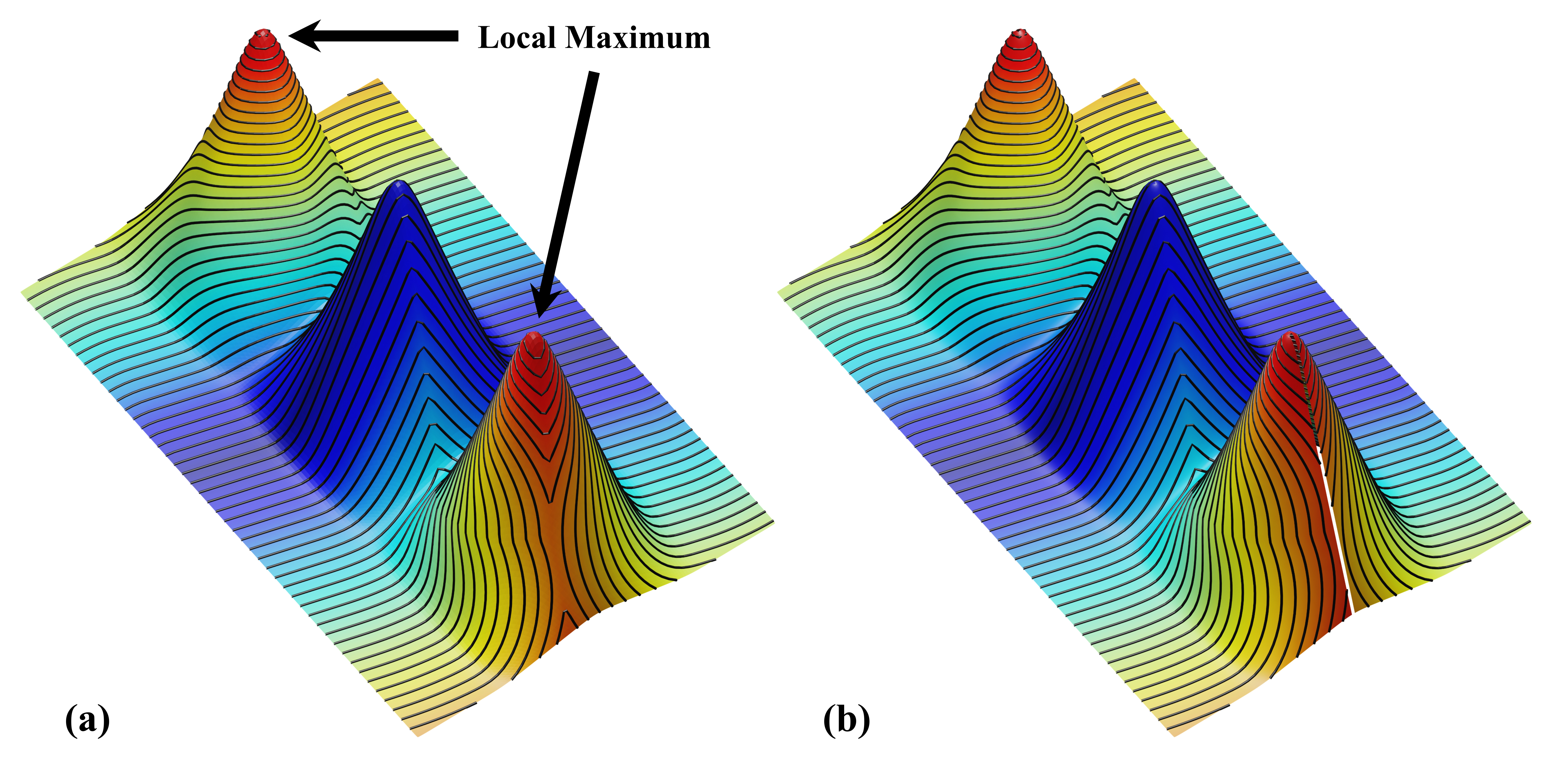}\vspace{-10pt}
\caption{\revnew{}{An example of failure case. (a) Local maximum will be generated when selecting the middle curves as source of the the geodesic distance field. On a field with local maximum located inside the surface patch, closed loops of isocurves will be formed in which case our algorithm of weaving map generation will fail. (b) Weaving map generation algorithm can only be applied after adding darts from the local maximum to place it on the boundary of a surface patch. This is how the Triple-Peak example shown in Figs.\ref{fig:FabResultsAll} and \ref{fig:ResGeo} was fabricated.}
}\label{fig:Limitation}
\end{figure}

There are also some limitation on the software aspect. Our current method of weaving-map generation heavily relies on selecting a good curve as source to generate geodesic distance-field. Different maps including failure cases can be generated when using different sources. 
\revnew{}{One failure scenario occurs when the local minimum or maximum appears after applying the geodesic distance field, leading to the generation of isocurves in closed loops for weft threads (see Fig.\ref{fig:Limitation}(a) for an example). To resolve this issue, we need to cut the given model by adding darts so that pushing the local minimum or maximum to the boundary of a surface patch to successfully generate the weaving map (as shown in Fig.\ref{fig:Limitation}(b)).} 
One possible future research is to automate the source selection while preventing the failure cases and reducing the distortion of stitch meshes constructed from the geodesic distance-field.

\section{Conclusion}
This paper presents a new computer-controlled 3D surface weaving method, which is the first approach that automates the fabrication of 3D freeform woven fabrics by using threads in non-traditional materials with high bending-stiffness. The major contributions of this work come from \rev{both}{the new process,} the mechatronic system design and the computational algorithm to enable the computer-controllable 3D surface weaving process. The performance of our technology has been tested and verified on a variety of 3D freeform models\rev{}{~and different applications}.

\section*{References}
\bibliographystyle{elsarticle-num}
\bibliography{reference}

\newpage

\section*{Appendix A: Detailed Schematics and PCB Layout} 
\noindent In this appendix, we provide the detailed schematics and PCB layout of our 3D surface weaving machine's control system. 

\begin{figure}[!h]
\centering 
\includegraphics[width=\linewidth]{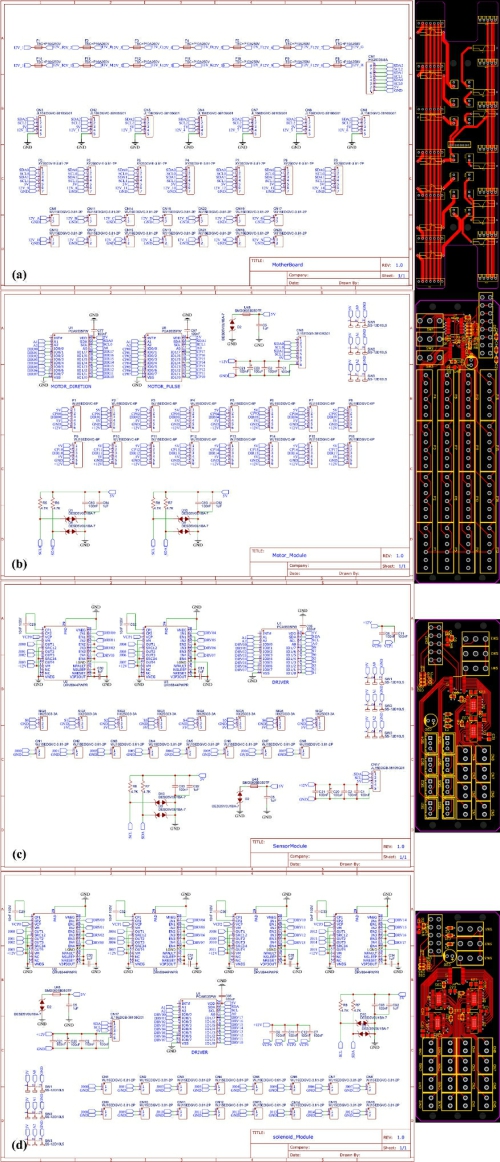}\vspace{-5pt}
\caption{The detailed schematics and PCB layouts for the motherboard (a), the module of motor (b), the module of sensor (c), and the module of solenoid (d).}
\label{fig:SchematicsPCB}
\end{figure}

\newpage
\noindent The whole control system consists of a motherboard, 7 motor modules, 6 solenoid modules, and one sensor module. 
See Fig.\ref{fig:SchematicsPCB} for the details.
The motherboard with 12V-100A as power input has 14 interfaces for different modules, which allows quick change and further module development. 
Besides of PCA9535 as the IO extension chip, the driver chip DRV8844 is employed as a power transformer in the solenoid modules. 

\section*{Appendix B: Example W-Code}
\noindent We list the example W-code of the back piece of the vest below in Fig.\ref{fig:Wcode}. The code is corresponding to the weaving map in coarse resolution that has been shown in Fig.8(d). 

\begin{figure}[!h]
\centering 
\includegraphics[width=\linewidth]{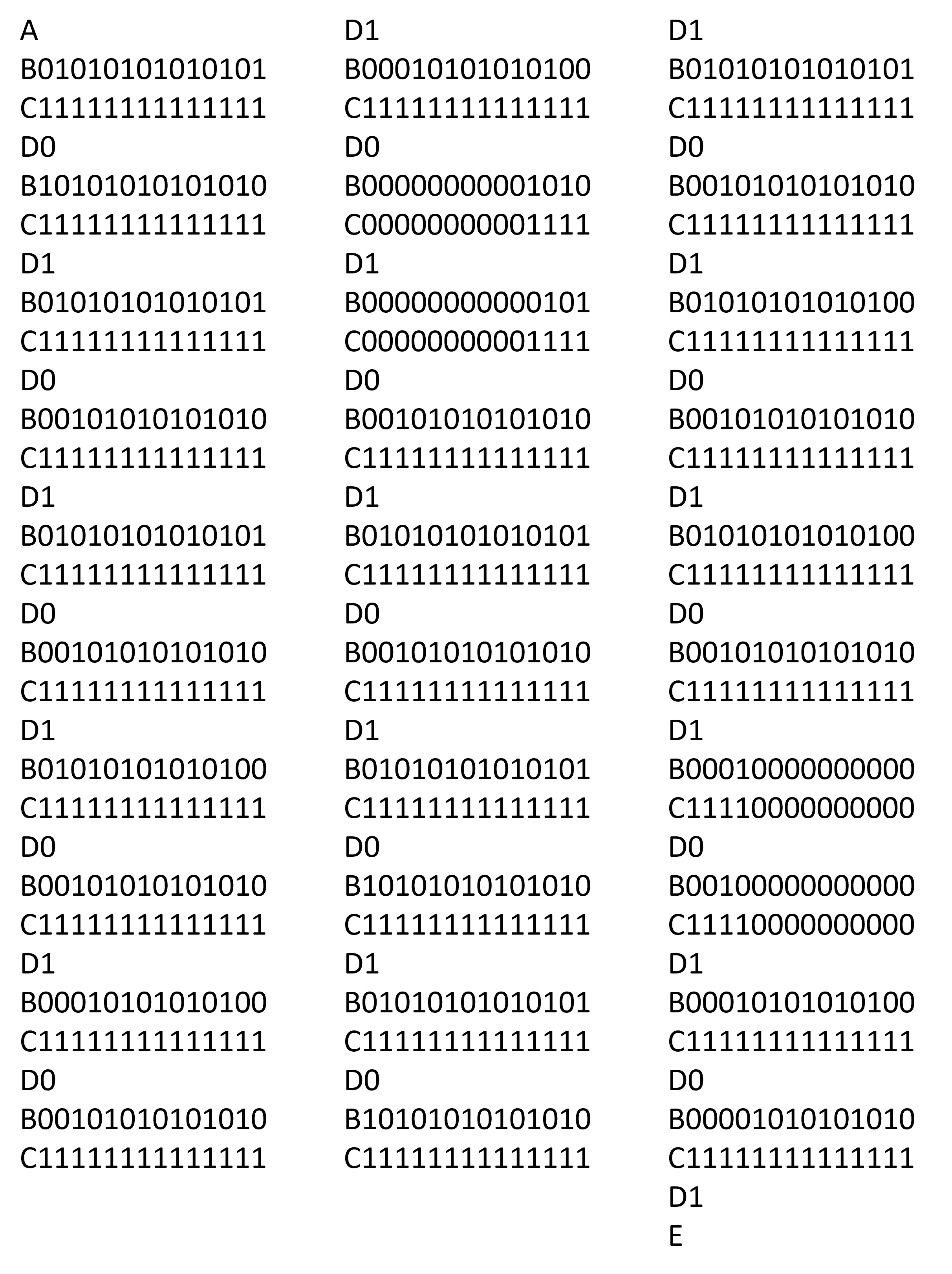}\vspace{-5pt}
\caption{The example W-code for the back piece of the vest (listed in three columns from left to right).}
\label{fig:Wcode}
\end{figure}

\end{document}